\documentclass[aps,prl,onecolumn,showpacs,preprintnumbers]{revtex4}
\usepackage{amsmath}
\usepackage{dcolumn}
\usepackage{bm}
\usepackage{subfigure}
\usepackage{amsfonts}
\usepackage{amssymb}
\usepackage{makeidx}
\usepackage{epsfig}
\usepackage{graphicx}

\begin{document}

\title[]{\textbf{On the conditions of validity of the Boltzmann equation and
Boltzmann H-theorem}}
\author{Massimo Tessarotto\thanks{%
email:M.Tessarotto@units.it; web-site: http://cmfd.units.it}}
\affiliation{Department of Mathematics and Geosciences, University of Trieste, Italy}
\author{Claudio Cremaschini}
\affiliation{Institute of Physics, Faculty of Philosophy and Science, Silesian University
in Opava, Bezru\v{c}ovo n\'{a}m.13, CZ-74601 Opava, Czech Republic}
\author{Marco Tessarotto}
\affiliation{Civil Protection Agency of Friuli Venezia-Giulia, Palmanova (Udine), Italy}
\date{\today }

\begin{abstract}
In this paper the problem is posed of the formulation of the so-called
\textquotedblleft ab initio\textquotedblright\ approach to the statistical
description of the Boltzmann-Sinai $N-$body classical dynamical system (CDS)
formed by identical smooth hard spheres. This amounts to introducing a
suitably-generalized version of the axioms of Classical Statistical
Mechanics. The latter involve a proper definition of the functional setting
for the $N-$body probability density function (PDF), so that it includes
also the case of the deterministic $N-$body PDF. In connection with this
issue, a further development concerns the introduction of modified collision
boundary conditions which differ from the usual ones adopted in previous
literature. Both features are proved to be consistent with the validity of
exact H-theorems for the $N-$body and $1-$body PDFs respectively.

Consequences of the axiomatic approach which concern the conditions of
validity of the Boltzmann kinetic equation and the Boltzmann H-theorem are
investigated. In particular, the role of the modified boundary conditions is
discussed. It is shown that both theorems fail in the case in which the $N-$%
body PDF is identified with the deterministic PDF. Finally, the issue of
applicability of the Zermelo and Loschmidt paradoxes to the
\textquotedblleft ab initio\textquotedblright\ approach presented here is
discussed.
\end{abstract}

\maketitle

%\tableofcontents

%\frontmatter

%\email{cremasch@sissa.it}

%\email{M.Tessarotto@cmfd.univ.trieste.it}

\section{Introduction}

In 1872 Ludwig Boltzmann published his famous paper \cite{Boltzmann1972} on
the equation bearing his name, which describes the statistical behavior of a
$N-$body classical dynamical system (CDS) formed by a large (i.e., with $%
N\gg 1$) ensemble of identical smooth hard spheres of constant diameter $%
\sigma $ ($S_{N}-$CDS). In his paper he proved, at the same time, also the
irreversibility property of the Boltzmann equation, later to become known in
the literature as the so-called Boltzmann H-theorem. The success of
Boltzmann's theory was initially slow and met with strong contemporary
critiques. Well-known in this respect are the objections raised by Loschmidt
and Zermelo (Loschmidt, 1876 \cite{Loschmidt1876} and Zermelo \cite%
{Zermelo-1,Zermelo-II}). These concern the apparent contradictions between
the Boltzmann equation and H-theorem with respect to the microscopic
reversibility property and the Poincar\`{e} recurrence theorem on the energy
surface (see also the replies given by Boltzmann in Refs.\cite%
{Boltzmann1896,Boltzmann11896b} and included also in his two-volume
text-book published in the same years \cite{Boltzmann1896c}). The latter
ones, in fact, are characteristic properties of the underlying CDS. Since
then, in the subsequent 140 years, the Boltzmann paper has come to be
acknowledged as one of the corner-stones of the kinetic theory of gases and
a popular subject of scientific research. In the course of time, a host of
investigations has been devoted to the subject. Some of these contributions
have been very influential both to the physical and mathematical
communities. A most significant example of this type is that due to Grad
(Grad, 1958 \cite{Grad}), who first attempted a first-principle derivation
of the Boltzmann equation based on an axiomatic approach to classical
statistical mechanics (CSM) relying on the $N-$body Liouville equations and
the related BBGK hierarchy for the $S_{N}-$CDS. For this purpose he
introduced a limiting process, known as Boltzmann-Grad limit, which
involves, in particular, the adoption of a suitable asymptotic ordering,
denoted as\emph{\ }rarefied-gas ordering, which is applicable only in the
case of rarefied gases (for a discussion of the concept see Refs.\cite%
{Grad,Lanford}). The same approach, which actually relies on appropriate
smoothness assumption for the relevant probability density functions (PDF),
was subsequently adopted in the literature by many other authors (see for
example Cercignani, Refs.\cite%
{Cercignani1969a,Cercignani1975,Cercignani1988,Cercignani2008}).

However, despite significant developments which concern primarily
mathematical properties of the theory, important aspects still remain
unsettled to date. These arise in particular both due to \textquotedblleft
ad hoc\textquotedblright\ physical assumptions invoked in the Boltzmann
original approach \cite{Boltzmann1972} and the asymptotic character of the
Boltzmann equation itself. Indeed, besides the requirement posed by the
Boltzmann-Grad asymptotic ordering, the Boltzmann equation requires the
validity of the Boltzmann \textquotedblleft
stosszahlansatz\textquotedblright\ condition. Thus, for example\ it is
well-known that it does not hold in the case in which the gas is locally
dense and undergoes strong density variations on the same scale \cite%
{Chapman-Cowling,Enskog}.

Another important issue concerns the choice of the functional setting for
the $N-$body probability density and the related collision boundary
conditions. The latter refer to the prescription of the outgoing $N-$body
PDF (after an arbitrary collision event) in terms of the corresponding
incoming PDF (before collision). In this sense, it is interesting to remark
that the need to modify the boundary conditions originally adopted by
Boltzmann was implicit in the approach proposed by Enskog \cite{Enskog} and
motivated by the treatment of dense gases. In such a case, in fact, the
finite size of interacting sphere must be taken into account (see also
related discussion in Chapman and Cowling, Ref.\cite{Chapman-Cowling}).

It is obvious that both issues are matters of principle for the proper
statistical treatment of real gases, both in the case of dense and rarefied
systems. The epitome example remains the classical dynamical system $S_{N}-$%
CDS originally proposed by Boltzmann himself and later thoroughly
investigated by several authors (which are summarized in Sinai \cite%
{Sinai1970,Sinai1989} and Anosov and Sinai \cite{Anosov-Sinai1967}), hereon
referred to as \emph{Boltzmann-Sinai CDS. }Its definition and basic
properties are recalled for completeness in the Appendix A. A consistent
solution of these issues can only be addressed in the framework of the
axiomatic formulation based on CSM. This type of treatment is denoted here
as\ \textquotedblleft \emph{ab initio}\textquotedblright\ approach to the
statistical description of the Boltzmann-Sinai CDS.

\subsection{Approaches to the kinetic statistical description}

It is well-known that for a set of classical identical particles, the
kinetic statistical description is realized via the construction of an
appropriate kinetic statistical equation for the $1-$body PDF. The latter is
here indicated as $\rho _{1}^{\left( N\right) }$ and is defined on the
phase--space $\Gamma _{1}\equiv \Omega _{1}\times U_{1}$, with $\Omega
_{1}\subseteq
%TCIMACRO{\U{211d} }%
%BeginExpansion
\mathbb{R}
%EndExpansion
^{3}$ and $U_{1}\equiv
%TCIMACRO{\U{211d} }%
%BeginExpansion
\mathbb{R}
%EndExpansion
^{3}$ denoting respectively the bounded $1-$body Euclidead configuration
space and the corresponding velocity space. The result can in principle be
equivalently achieved following different routes. The first one, due to
Boltzmann himself (Boltzmann, 1872 \cite{Boltzmann1972}), follows directly
from the analysis of the $\Gamma _{1}-$phase--space dynamics of $\rho
_{1}^{\left( N\right) }$. The second approach, which is due to Grad (Grad,
1958 \cite{Grad}), relies instead on the $\Gamma _{N}-$phase--space
differential Liouville equation which is assumed to hold for the $N-$body
PDF $\rho ^{\left( N\right) }$ in the sub-set of $\Gamma _{N}$ in which no
interactions occur. In particular, here $\Gamma
_{N}=\prod\limits_{i=1,N}\Gamma _{1(i)}$, with $\Gamma _{1(i)}=\Omega
_{1(i)}\times U_{1(i)}$ being the $i-$th particle phase-space, while $\Omega
_{1(i)}\subseteq
%TCIMACRO{\U{211d} }%
%BeginExpansion
\mathbb{R}
%EndExpansion
^{3}$ and $U_{1(i)}\equiv
%TCIMACRO{\U{211d} }%
%BeginExpansion
\mathbb{R}
%EndExpansion
^{3}$ are the corresponding Euclidean configuration and velocity spaces for
the same particle. This involves in principle the construction of the whole
BBGKY hierarchy for the set of reduced $s-$body probability densities $%
\left\{ \rho _{s}^{\left( N\right) },s=1,N-1\right\} $. In the second case,
according to Grad (see also Cercignani \cite{Cercignani1975,Cercignani1988}%
), the collision boundary conditions for $\rho ^{\left( N\right) }$ are
taken to be of the form:%
\begin{equation}
\rho ^{(+)\left( N\right) }(\mathbf{x}^{\left( +\right) }(t_{i}),t_{i})=\rho
^{(-)\left( N\right) }(\mathbf{x}^{\left( -\right) }(t_{i}),t_{i}),
\label{PEF CONSERVATION}
\end{equation}%
with $t_{i}\in \left\{ t_{i}\right\} \equiv \left\{ t_{i},i\in
%TCIMACRO{\U{2115} }%
%BeginExpansion
\mathbb{N}
%EndExpansion
\right\} $ being an arbitrary collision time belonging to the continuous
time interval $I\subseteq
%TCIMACRO{\U{211d} }%
%BeginExpansion
\mathbb{R}
%EndExpansion
$. Eq.(\ref{PEF CONSERVATION}) is referred to as \emph{PDF-conserving
collisional boundary condition}. This prescription is also consistent with
the original Boltzmann approach \cite{Boltzmann1972}. Here,\ $\rho
^{(+)\left( N\right) }(\mathbf{x}^{\left( +\right) }(t_{i}),t_{i})$ and $%
\rho ^{(-)\left( N\right) }(\mathbf{x}^{\left( -\right) }(t_{i}),t_{i})$
identify respectively the $N-$body PDFs after and before collision, namely%
\begin{equation}
\rho ^{(+)\left( N\right) }(\mathbf{x}^{\left( +\right)
}(t_{i}),t_{i})=\lim_{t\rightarrow t_{i}^{\left( +\right) }}\rho ^{\left(
N\right) }(\mathbf{x}(t),t).  \label{PDF AFTER COLLISION}
\end{equation}%
\begin{equation}
\rho ^{(-)\left( N\right) }(\mathbf{x}^{\left( -\right)
}(t_{i}),t_{i})=\lim_{t\rightarrow t_{i}^{\left( -\right) }}\rho ^{\left(
N\right) }(\mathbf{x}(t),t),  \label{PDF BEFOR COLLISION}
\end{equation}%
Nevertheless, this type of boundary condition, which is customarily adopted
in the construction of the Boltzmann equation, is generally violated at
least in the following cases:

A) By the deterministic $N-$body PDF (or \textquotedblleft certainty
PDF\textquotedblright\ according to Ref.\cite{Cercignani1969a}), i.e., the $%
N-$body Dirac delta $\rho _{H}^{\left( N\right) }(\mathbf{x},t)=\delta
\left( \mathbf{x}-\mathbf{x}(t)\right) $ (see related discussion below, and
in particular Eqs.(\ref{MULTIDIMENSIONAL DIRAC DELTA})-(\ref{DIRAC DELTA-
FACTORIZED FORM})). Therefore, on the support of $\rho _{H}^{\left( N\right)
}(\mathbf{x},t)$, one obtains%
\begin{equation}
\rho _{H}^{\left( N\right) }(\mathbf{x},t)=\rho _{H}^{\left( N\right) }(%
\mathbf{x}\left( t\right) ,t).
\end{equation}%
Then, in analogy to Eqs.(\ref{PDF BEFOR COLLISION})-(\ref{PDF AFTER
COLLISION}), denoting $\rho _{H}^{\left( -\right) \left( N\right) }(\mathbf{x%
}^{\left( -\right) }(t_{i}),t_{i})=\lim_{t\rightarrow t_{i}^{\left( -\right)
}}\rho _{H}^{\left( N\right) }(\mathbf{x}(t),t)$, and $\rho _{H}^{\left(
+\right) \left( N\right) }(\mathbf{x}^{\left( +\right)
}(t_{i}),t_{i})=\lim_{t\rightarrow t_{i}^{\left( +\right) }}\rho
_{H}^{\left( N\right) }(\mathbf{x}(t),t)$, it follows that on its support, $%
\rho _{H}^{\left( N\right) }(\mathbf{x},t)$ must satisfy the boundary
conditions%
\begin{equation}
\rho _{H}^{(+)\left( N\right) }(\mathbf{x}^{\left( +\right)
}(t_{i}),t_{i})=\rho _{H}^{(-)\left( N\right) }(\mathbf{x}^{\left( +\right)
}(t_{i}),t_{i}).  \label{BC dirac}
\end{equation}

B) By the factorized $2-$body PDF%
\begin{equation}
\rho _{2}^{\left( N\right) }(\mathbf{x}_{1},\mathbf{x}_{2},t)=\rho
_{1}^{\left( N\right) }(\mathbf{x}_{1},t)\rho _{1}^{\left( N\right) }(%
\mathbf{x}_{2},t),  \label{STOSS-ANSATZ}
\end{equation}%
which is introduced in the Boltzmann collision operator\ after invoking the
so-called Boltzmann \textquotedblleft stosszahlansatz\textquotedblright\
assumption (see Eq.(\ref{C2}) below). Indeed, it is obvious that $\rho
_{2}^{\left( N\right) }(\mathbf{x}_{1},\mathbf{x}_{2},t)$ cannot generally
fulfill the boundary condition of the type (\ref{PEF CONSERVATION}), because
in such a case \textit{\textquotedblleft there would be no effect of the
collisions on the time evolution\textquotedblright\ }of\textit{\ }$\rho
_{1}^{\left( N\right) }(\mathbf{x}_{1},t)$ (Cercignani \cite{Cercignani2008}%
). Rather, in analogy to (\ref{BC dirac}), it is natural to impose on $\rho
_{2}^{\left( N\right) }(\mathbf{x}_{1},\mathbf{x}_{2},t)$ modified boundary
conditions of the form%
\begin{equation}
\rho _{2}^{(+)\left( N\right) }(\mathbf{x}_{1}^{\left( +\right) }(t_{i}),%
\mathbf{x}_{2}^{\left( +\right) }(t_{i}),t_{i})=\rho _{2}^{(-)\left(
N\right) }(\mathbf{x}_{1}^{\left( +\right) }(t_{i}),\mathbf{x}_{2}^{\left(
+\right) }(t_{i}),t_{i}).
\end{equation}

The two examples already suggest that the boundary conditions indicated
above by Eq.(\ref{PEF CONSERVATION}) can become unphysical. This means that
the PDF-conserving boundary condition should be replaced by a suitable new
one consistent with the treatment of cases A) and B).

In order to solve the problem and formulate in a systematic way the
statistical description of CDSs of this type, a third independent approach
is adopted here, which is based on the axiomatic formulation of CSM for the $%
N-$body $S_{N}-$CDS. For definiteness, we assume that the same CDS is
prescribed by means of a bijection onto the phase space $\Gamma _{N}$ of the
form (see also Appendix A):
\begin{equation}
T_{t_{o},t}:\mathbf{x}(t_{o})\equiv \mathbf{x}_{o}\rightarrow \mathbf{x}%
(t)\equiv \mathbf{\chi }(\mathbf{x}_{o},t_{o},t)\equiv T_{t_{o},t}\mathbf{x}%
_{o},  \label{DS}
\end{equation}%
with inverse transformation%
\begin{equation}
T_{t,t_{o}}:\mathbf{x}\equiv \mathbf{x}(t)\rightarrow \mathbf{x}(t_{o})=%
\mathbf{x}_{o}=\mathbf{\chi }(\mathbf{x},t,t_{o})\equiv T_{t,t_{o}}\mathbf{x}%
_{o}.  \label{DS-1}
\end{equation}

Then the statistical description for the $S_{N}-$CDS is realized by invoking
the following axioms of CSM:

\begin{itemize}
\item \emph{Axiom \#1 of probability - }This is based on the introduction of
a probability density $\rho ^{\left( N\right) }(t)\equiv \rho ^{\left(
N\right) }(\mathbf{x},t)$ on the $N-$body phase-space $\Gamma _{N}$. As a
consequence, the probability of an arbitrary subset $A\equiv A(t)$ of $%
\Gamma _{N}$ is given by%
\begin{equation}
P\left\{ A(t)\right\} =\int\limits_{A}d\mathbf{x}\rho ^{\left( N\right) }(%
\mathbf{x},t).  \label{ax1}
\end{equation}%
Here, by assumption $A(t)$ and $\rho ^{\left( N\right) }(\mathbf{x},t)$ are
such that: 1) $A(t)$ belongs to an appropriate family of subsets of $\Gamma
_{N}$ denoted as $K(\Gamma _{N})$; 2) $\rho ^{\left( N\right) }(t)$ is
required to belong to a suitable functional space $\left\{ \rho ^{\left(
N\right) }(t)\right\} $ which necessarily must include the deterministic $N-$%
body PDF $\rho _{H}^{\left( N\right) }(t)\equiv \rho _{H}^{\left( N\right) }(%
\mathbf{x},t)$ (see also related discussion in Appendix B); 3) all the PDFs
belonging to the functional space $\left\{ \rho ^{\left( N\right)
}(t)\right\} $ must fulfill identically the property of time-reversal
invariance:%
\begin{equation}
\rho ^{\left( N\right) }(\mathbf{r},\mathbf{v},\tau )=\rho ^{\left( N\right)
}(\mathbf{r},-\mathbf{v},-\tau ),  \label{treversal}
\end{equation}%
where $\tau =t-t_{1}\in I$ and $t_{1}\in I$ is an arbitrary fixed reference
time, for all $\mathbf{x}=(\mathbf{r},\mathbf{v})\equiv \mathbf{x}\left(
t\right) $.

\item \emph{Axiom \#2 of probability conservation - }The axiom requires for
arbitrary $t_{o},t\in I$\emph{\ }the probability conservation law%
\begin{equation}
P\left\{ A(t)\right\} =P\left\{ A(t_{o})\right\} ,
\label{PROBABIULITY CONSERVATION LAW}
\end{equation}%
which must apply for an arbitrary ensemble $A(t_{o})$ and its image $A(t)$
determined by the CDS (\ref{DS})-(\ref{DS-1}), both belonging to a suitable
family of sets $K(\Gamma _{N})$.

\item \emph{Axiom \#3 of entropy maximization - }This requires that the
initial PDF $\rho ^{\left( N\right) }(t_{o})\equiv \rho ^{\left( N\right) }(%
\mathbf{x},t_{o})$ maximizes at $t=t_{o\text{ }}$a suitably-defined
statistical entropy $S_{N}\left( \rho ^{\left( N\right) }(t_{o})\right) $,
to be identified with the $N-$body Boltzmann-Shannon entropy associated with
the same PDF (see definition in Appendix B and Refs.\cite%
{Shannon,Jaynes1957a,Jaynes1957b}). A particular case must include the
deterministic PDF $\rho _{H}^{\left( N\right) }(t_{o})$.
\end{itemize}

Regarding Axiom \#1 in particular, it is important to remark that the
requirement of existence of the deterministic PDF $\rho _{H}^{\left(
N\right) }(\mathbf{x},t)$ must actually apply to arbitrary CDSs (not only to
the case of $S_{N}-$CDS). This condition, in fact, corresponds to the
deterministic description for the same CDSs.

In the following only the first two axioms will be actually used.
Nevertheless the appropriate definition of $S_{N}\left( \rho ^{\left(
N\right) }(t)\right) $ is still needed for comparisons with previous
approaches to kinetic theory. This involves the appropriate prescription of
the functional setting for the $N-$body dynamical system, i.e., besides the
specification of $K(\Gamma _{N})$, the definition of the functional class $%
\left\{ \rho ^{\left( N\right) }(t)\right\} $. We remark that in difference
with the approach by Grad, where the differential Liouville equation is
postulated from the start, the \textquotedblleft ab
initio\textquotedblright\ approach developed here and based on the axioms of
CSM allows one to suitably prescribe both the set $K(\Gamma _{N})$ and $%
\left\{ \rho ^{\left( N\right) }(t)\right\} $.

\section{Goals of the paper}

In this paper we set up a general framework for the statistical treatment of
the Boltzmann-Sinai dynamical system based on the\ \textquotedblleft ab
initio\textquotedblright\ approach outlined above. Based exclusively on the
physical prescriptions dictated by CSM, our goal is to show that the
following conclusions apply:

\begin{enumerate}
\item[\emph{Claim \#1}] \emph{- Functional class }$\left\{ \rho ^{\left(
N\right) }(t)\right\} $\emph{:} $\left\{ \rho ^{\left( N\right) }(t)\right\}
$ can be uniquely defined in such a way to satisfy the axioms of CSM for all
$\rho ^{\left( N\right) }(\mathbf{x},t)\in \left\{ \rho ^{\left( N\right)
}(t)\right\} $, including distributions, which may violate the
PDF-conserving boundary condition (\ref{PEF CONSERVATION}). In particular it
is proved that the functional class $\left\{ \rho ^{\left( N\right)
}(t)\right\} $ must generally include \emph{partially deterministic} PDFs of
the form%
\begin{equation}
\rho ^{\left( N\right) }(\mathbf{x},t)=\delta \left( \mathbf{f}(\mathbf{x}%
,t)\right) w^{\left( N\right) }(\mathbf{x},t)\equiv \rho _{d}^{\left(
N\right) }(\mathbf{x},t),  \label{PDF-1}
\end{equation}%
with $\delta \left( \mathbf{f}(\mathbf{x},t)\right) $, $\mathbf{f}(\mathbf{x}%
,t)$ and $w^{\left( N\right) }(\mathbf{x},t)$ denoting respectively a
multi-dimensional Dirac delta, a suitable smooth real vector function and a
strictly positive smooth and summable function. In particular, these include
the $N\emph{-}$\emph{body deterministic PDF} $\rho _{H}^{\left( N\right) }(%
\mathbf{x},t)$ defined above. The corresponding $1-$body PDF obtained by
integrating $\rho _{H}^{\left( N\right) }(\mathbf{x},t)$ on the subset of
phase-space $\Gamma _{N}$, $\Gamma _{2}\times \Gamma _{3}...\times \Gamma
_{N}$, therefore necessarily coincides with the $1$\emph{-body deterministic
PDF:}%
\begin{equation}
\rho _{1}^{\left( N\right) }(\mathbf{x}_{1},t)=\delta \left( \mathbf{x}_{1}-%
\mathbf{x}_{1}\left( t\right) \right) \equiv \rho _{H1}^{\left( N\right) }(%
\mathbf{x}_{1},t).  \label{1-body deterministic PDF}
\end{equation}%
Here the notation is standard. Thus, $(\mathbf{x}_{i},t)$ for $i=1,N$
denotes the extended $i-$particle phase-state, with $\mathbf{x}_{i}=\left(
\mathbf{r}_{i},\mathbf{v}_{i}\right) $ being the Newtonian state which spans
the phase-space $\Gamma _{1(i)}$. In addition, denoting by $\mathbf{x}%
=\left\{ \mathbf{x}_{1},..,\mathbf{x}_{N}\right\} $ the $N-$body Newtonian
state, for all $i=1,N,$ $\mathbf{x}_{i}\left( t\right) $ is determined
uniquely by the Newtonian CDS:%
\begin{equation}
\mathbf{x}_{_{0}}\equiv \left\{ \mathbf{x}_{1}\left( t_{o}\right) ,..,%
\mathbf{x}_{N}\left( t_{o}\right) \right\} \rightarrow \mathbf{x}\left(
t\right) \equiv \left\{ \mathbf{x}_{1}\left( t\right) ,..,\mathbf{x}%
_{N}\left( t\right) \right\} .  \label{NEWTONIAN CDS}
\end{equation}

\item[\emph{Claim \#2}] \emph{- Modified collision boundary condition
(MCBC):\ }Thanks to Axiom \#1 it is shown (see Lemma to THM.1) that for an
arbitrary PDF $\rho ^{\left( N\right) }(\mathbf{x},t)\in \left\{ \rho
^{\left( N\right) }(t)\right\} $, with $(\mathbf{x},t)\in \Gamma _{N}\times
I $, and arbitrary collision times $t_{i}\in \left\{ t_{i}\right\} \equiv
\left\{ t_{i},i\in
%TCIMACRO{\U{2115} }%
%BeginExpansion
\mathbb{N}
%EndExpansion
\right\} $, modified collision boundary conditions (MCBC) necessarily apply.
In Lagrangian form they are of the type
\begin{equation}
\rho ^{(+)\left( N\right) }(\mathbf{x}^{\left( +\right) }(t_{i}),t_{i})=\rho
^{\left( -\right) \left( N\right) }(\mathbf{x}^{(+)}(t_{i}),t_{i}),
\label{BC-1}
\end{equation}%
with $\mathbf{x}^{\left( +\right) }(t_{i})$ denoting the post-collision
state originating from $\mathbf{x}^{(-)}(t_{i})$ at a generic collision time
$t_{i}\in \left\{ t_{i}\right\} $ of a Lagrangian trajectory generated by $%
S_{N}-$CDS. Here, $\rho ^{(+)\left( N\right) }(\mathbf{x}^{\left( +\right)
}(t_{i}),t_{i})$ identifies the $N-$body PDF after collision, i.e., the
limit (\ref{PDF AFTER COLLISION}), while $\rho ^{\left( -\right) \left(
N\right) }(\mathbf{x}^{(-)}(t_{i}),t_{i})$ and $\mathbf{x}^{(-)}(t_{i})$
denote respectively the corresponding incoming PDF and the state before
collision. Then, if $\rho ^{\left( N\right) }(\mathbf{x}(t),t)$ is
left-continuous at all collision times $t_{i}$, the previous equation can be
replaced with%
\begin{equation}
\rho ^{(+)\left( N\right) }(\mathbf{x}^{\left( +\right) }(t_{i}),t_{i})=\rho
^{\left( N\right) }(\mathbf{x}^{(+)}(t_{i}),t_{i}).  \label{BC-2}
\end{equation}%
The corresponding Eulerian form of MCBC is then given by%
\begin{equation}
\rho ^{(+)\left( N\right) }(\mathbf{x}^{\left( +\right) },t)=\rho ^{\left(
N\right) }(\mathbf{x}^{(+)},t),  \label{BC-2-1}
\end{equation}%
with $\mathbf{x}$ and $\mathbf{x}^{\left( +\right) }$ denoting a suitable
set of pre- and post-collision states. From Eqs.(\ref{BC-2}) and (\ref%
{BC-2-1}) it follows that, in contrast to Eq.(\ref{PEF CONSERVATION}), for
an arbitrary $\rho ^{\left( N\right) }(\mathbf{x},t)\in \left\{ \rho
^{\left( N\right) }(t)\right\} $ and arbitrary collision time $t_{i}\in
\left\{ t_{i}\right\} $, the $N-$body probability density is generally not
conserved during collision events.

\item[\emph{Claim \#3}] \emph{- Probability conservation in }$K(\Gamma _{N})$%
\emph{:\ }The family\emph{\ }$K(\Gamma _{N})$ can be uniquely defined in
such a way to satisfy for all $A(t)\in K(\Gamma _{N})$ the Axiom \#2 of
probability conservation (THM.1).

\item[\emph{Claim \#4}] \emph{- H-theorems for the }$N-$\emph{body and }$1-$%
\emph{body PDFs: }At all times $t\in I$, an arbitrary $N\emph{-}$body PDF%
\emph{\ }$\rho ^{\left( N\right) }(\mathbf{x},t)\in \left\{ \rho ^{\left(
N\right) }(t)\right\} $ satisfies the constant H-theorem:%
\begin{equation}
\frac{\partial }{\partial t}S_{N}\left( \rho ^{\left( N\right) }(t)\right) =0
\label{constant-H-theorem N-body}
\end{equation}%
(THM.2). In particular, when $\rho ^{\left( N\right) }(\mathbf{x},t)$
coincides with $\rho _{H}^{\left( N\right) }(\mathbf{x},t)$ it follows
identically that%
\begin{equation}
S_{N}\left( \rho _{H}^{\left( N\right) }(t)\right) =0.
\end{equation}%
As a consequence, the corresponding $1-$body PDFs\emph{\ }$\rho _{1}^{\left(
N\right) }(\mathbf{x}_{1},t)$ and $\rho _{H1}^{\left( N\right) }(\mathbf{x}%
_{1},t)$ satisfy respectively a weak H-theorem of the form%
\begin{equation}
\frac{\partial }{\partial t}S_{1}\left( \rho _{1}^{\left( N\right)
}(t)\right) \geq 0,  \label{WEAK H-THEOREM-0}
\end{equation}%
and the constant H-theorem%
\begin{equation}
\frac{\partial }{\partial t}S_{1}\left( \rho _{H1}^{\left( N\right)
}(t)\right) =0,  \label{CONSTANT H-THEOREM -1-body}
\end{equation}%
with $S_{1}\left( \rho _{1}^{\left( N\right) }(t)\right) $ and $S_{1}\left(
\rho _{H1}^{\left( N\right) }(t)\right) $ denoting the corresponding $1-$%
body BS entropies (THM.3).
\end{enumerate}

The subsequent sections are devoted to the proof of the above statements. As
an application, we intend to discuss here (with particular reference to
Claims \#2-\#4) the relationship of the present theory respectively with the
Boltzmann kinetic equation and Boltzmann H-theorem \cite{Boltzmann1972}. The
issue is relevant in order to ascertain: a) their conditions of validity in
the present context; b) whether there are particular realizations of the $N-$%
body and corresponding $1-$body PDFs for which both the Boltzmann equation
as well as the Boltzmann H-theorem may possibly be violated; c) the possible
role of Loschmidt and Zermelo objections; d) the physical implications and
the reasons of their (possible) failure.\ In this connection, in the
following we intend to prove that:

\begin{enumerate}
\item[\emph{Claim \#5}] \emph{- Condition of validity of the Boltzmann
equation:} The\emph{\ }Boltzmann equation\ does not admit as a particular
solution the deterministic $1-$body\ PDF $\rho _{H1}^{\left( N\right) }(%
\mathbf{x}_{1},t)$ (THM.4).

\item[\emph{Claim \#6}] \emph{- Condition of validity the Boltzmann
H-theorem: }The\emph{\ }Boltzmann H-theorem is not fulfilled by the
deterministic $1-$body\ PDF $\rho _{H1}^{\left( N\right) }(\mathbf{x}_{1},t)$
(THM.5).

\item[\emph{Claim \#7}] \emph{- Condition of consistency with the
microscopic reversibility of }$S_{N}-$\emph{CDS (Loschmidt paradox }\cite%
{Loschmidt1876}\emph{):}\textbf{\ }Microscopic reversibility is not at
variance with the validity of a weak H-theorem for the BS entropy associated
with the $1-$body PDF $S_{1}\left( \rho _{1}^{\left( N\right) }(t)\right) $
(see related discussion in Section 6).

\item[\emph{Claim \#8}] \emph{- Condition of consistency with the Poincar%
\`{e} recurrence theorem (Zermelo objection }\cite{Zermelo-1,Zermelo-II}%
\emph{):}\textbf{\ }the modified boundary condition introduced here in Eq.(%
\ref{BC-1}) is shown to be in agreement with the properties of $S_{N}-$CDS
and in particular with the Poincar\`{e} recurrence theorem (see again
Section 6).
\end{enumerate}

\bigskip

\section{The \textquotedblleft ab initio\textquotedblright\ approach: Claims
\#1-\#3}

In order to develop rigorously the \textquotedblleft ab
initio\textquotedblright\ approach for the $S_{N}-$CDS we refer to the
mathematical preliminaries in Appendix B. Upon invoking the axiom of
probability (Axiom \#1), these tools permit us to prescribe the functional
class $\left\{ \rho ^{\left( N\right) }(t)\right\} $. As a basic consequence
it follows that $\rho ^{\left( N\right) }(\mathbf{x},t)$ can be realized
either in terms of suitably smooth ordinary functions (stochastic PDFs) or
by appropriate distributions to be identified with the $N-$body
deterministic and partially deterministic PDFs (see definitions in Appendix
B).\ In particular, the $N-$body deterministic PDF is shown to be endowed
with a vanishing $N-$body Boltzmann-Shannon entropy for which a constant
H-theorem holds identically.

Let us formulate precisely the axioms of CSM. For a generic CDS (and hence
in particular for the Newtonian $S_{N}-$CDS), the first axiom goes as
follows:

\begin{enumerate}
\item[\emph{Axiom \#1}] \emph{Axiom of probability density on }$\Gamma _{N}$%
\emph{.} \emph{For any subset }$A\equiv A(t)$\emph{\ of }$\Gamma _{N}$\emph{%
\ and for all }$t\in I$\emph{, the probability of }$A(t)$\emph{\ is
prescribed according to Eq.(\ref{ax1}) in terms of }$\rho ^{\left( N\right)
}(\mathbf{x},t)$\emph{\ (i.e., the }$N-$\emph{body PDF on }$\Gamma _{N}$%
\emph{). The PDF is prescribed in such a way that:}

\item[1a] $\rho ^{\left( N\right) }(\mathbf{x},t)$\emph{\ is defined for all
}$(\mathbf{x},t)$ $\in \Gamma _{N}\times I.$

\item[1b] \emph{The PDF }$\rho ^{\left( N\right) }(\mathbf{x},t)$ \emph{can
be identified either with a stochastic, deterministic or
partially-deterministic PDF (see Appendix B).}

\item[1c] \emph{The initial }$N-$\emph{body PDF}$\rho ^{\left( N\right)
}(t_{o})\equiv \rho ^{\left( N\right) }(\mathbf{x}_{o},t_{o})$\emph{\
belongs to an appropriately-defined functional class }$\left\{ \rho ^{\left(
N\right) }(t_{o})\right\} .$\emph{\ This uniquely determines also the
functional class }$\left\{ \rho ^{\left( N\right) }(t)\right\} $ \emph{of
the time-evolved PDF} $\rho ^{\left( N\right) }(t)\equiv \rho ^{\left(
N\right) }(\mathbf{x},t).$ \emph{In particular, this requires that, if} $%
\rho ^{\left( N\right) }(t_{o})$ \emph{is deterministic, it remains so also
for all }$t\in I$.

\item[1d] $\rho ^{\left( N\right) }(\mathbf{x},t)\in \left\{ \rho ^{\left(
N\right) }(t)\right\} $\emph{\ must fulfill identically the property of
time-reversal invariance expressed by Eq.(\ref{treversal}).}

\item[1e] \emph{For all }$\rho ^{\left( N\right) }(\mathbf{x},t)\in \left\{
\rho ^{\left( N\right) }(t)\right\} $\emph{\ and for a suitable choice of
the weight-function }$G(\mathbf{x},t)$\emph{,} \emph{it is assumed that the
ensemble-averages}%
\begin{equation}
\left\langle G(\mathbf{x},t)\right\rangle =\int\limits_{\Gamma _{N}}d\mathbf{%
x}G(\mathbf{x},t)\rho ^{\left( N\right) }(\mathbf{x},t)
\label{ENSEMBLE-AVERAGE}
\end{equation}%
\emph{exist.}

\item[1f] \emph{The weight function} $G(\mathbf{x},t)$ \emph{as well as the
state} $\mathbf{x}(t)\in \Gamma _{N}$ \emph{can be either stochastic or
deterministic depending on the choice of the }$N-$\emph{body PDF.}
\end{enumerate}

Fundamental consequences of the Axiom \#1 are that:

\begin{itemize}
\item For all $t\in I$, the $N-$body deterministic PDF $\rho _{H}^{\left(
N\right) }(\mathbf{x},t)$ belongs to the functional class $\left\{ \rho
^{\left( N\right) }(t)\right\} $.

\item By construction, the $N-$body deterministic PDF $\rho _{H}^{\left(
N\right) }(\mathbf{x},t)$ is characterized by an identically vanishing\emph{%
\ }Boltzmann-Shannon entropy and also a vanishing corresponding entropy
production rate (see Eqs.(\ref{ZERO N-BODY ENTROPY}) and (\ref{CONSTANT
N-BODY ENTROPY}) in Appendix B).

\item The previous statements are consistent with Claim \#1.
\end{itemize}

\bigskip

Let us now address the issue related to the determination of the collision
boundary condition\emph{\ }for $\rho ^{\left( N\right) }(\mathbf{x},t)$
(Claim \#2). This is expressed by the Lemma:

\textbf{LEMMA to THM.1 - Lagrangian and Eulerian MCBC}

\emph{The following proposition holds: if }$\rho ^{\left( N\right) }(\mathbf{%
x},t)$ \emph{is an arbitrary PDF belonging to the functional class }$\left\{
\rho ^{\left( N\right) }(t)\right\} $\emph{, then it must necessarily
satisfy the modified Lagrangian boundary conditions (\ref{BC-1}). If }$\rho
^{\left( N\right) }(\mathbf{x}(t),t)$ \emph{is left-continuous at all }$%
t_{i}\in \left\{ t_{i}\right\} $\emph{\ then Eq.(\ref{BC-1}) is replaced
with Eq.(\ref{BC-2}). The corresponding Eulerian form of the boundary
conditions is then given by Eq.(\ref{BC-2-1}).}

\emph{Proof} - In fact, due to Axiom \#1, the boundary conditions must be
satisfied by the whole functional set $\left\{ \rho ^{\left( N\right)
}(t)\right\} $. Therefore, since the deterministic PDF belongs necessarily
to $\left\{ \rho ^{\left( N\right) }(t)\right\} $ and satisfies the boundary
conditions (\ref{BC dirac}), the proof of Eq.(\ref{BC-1}) follows.
Furthermore, representing (\ref{BC-1}) in terms of the analytic
pre-collision continuation of the CDS, i.e., the $S_{N}^{(-)}-$CDS (see
Appendix A, subsection 2), i.e., letting $\mathbf{x}^{(-)}(t_{i})=\mathbf{x}%
(t_{i})$ and assuming $\rho ^{\left( N\right) }(\mathbf{x}(t),t)$ to be
left-continuous at $t_{i}$, it follows that the previous equation can be
replaced with Eq.(\ref{BC-2}). The Eulerian form of the MCBC corresponding
to the last equation is then obtained by formally replacing $%
t_{i}\rightarrow t$ while letting $\mathbf{x}^{\left( -\right)
}(t_{i})\equiv \mathbf{x}$ and $\mathbf{x}^{\left( +\right) }(t_{i})\equiv
\mathbf{x}^{\left( +\right) }$, with $\mathbf{x}$ and $\mathbf{x}^{\left(
+\right) }$ denoting a suitable set of pre- and post-collision states.
Notice that their precise definition depends on the type of collision to be
considered (see Appendix A). This recovers the Eulerian form of the MCBC
given by Eq.(\ref{BC-2-1}).

\textbf{Q.E.D.}

The statement proves the validity of Claim \#2. The proof of the Lemma
applies manifestly also in the case in which arbitrary multiple collisions
are taken into account (see definition in Appendix A).

Let us now analyze the consequences of the Lemma and of the axiom of
probability (Axiom \#2). The following statement holds.

\textbf{THM.1 - Integral and differential Liouville equations for }$S_{N}-$%
\textbf{CDS}

\emph{Given validity of Lemma 1 to THM.1, let us assume that:}

\emph{1) The }$N-$\emph{body PDF for }$S_{N}-$\emph{CDS }$\rho ^{\left(
N\right) }(\mathbf{x},t)$ \emph{is a stochastic PDF;}

\emph{2)} $K(\Gamma _{N})$ \emph{includes only subsets of }$\Gamma _{N}$%
\emph{\ which are permutation symmetric, i.e., invariant with respect to
arbitrary permutations of the states of colliding particles.}

\emph{Then, it follows necessarily that:}

$T1_{1})$ \emph{If for }$i\in
%TCIMACRO{\U{2115} }%
%BeginExpansion
\mathbb{N}
%EndExpansion
,$ $I_{i}=\left] t_{i},t_{i+1}\right[ $ \emph{is an arbitrary open interval
between two consecutive collision times }$t_{i},t_{i+1}\in \left\{
t_{i}\right\} $\emph{, }$\rho ^{\left( N\right) }(\mathbf{x},t)$ \emph{%
satisfies the integral Liouville equation}%
\begin{equation}
\rho ^{\left( N\right) }(\mathbf{x}(t),t)=\rho ^{\left( N\right) }(\mathbf{x}%
\left( t_{o}\right) ,t_{o})  \label{INTEGRAL LIOUVILLE EQ}
\end{equation}%
\emph{for arbitrary }$t$ \emph{and} $t_{o}$ \emph{belonging to the \emph{%
same open time interval} }$I_{i}\emph{\ }$\emph{and for an arbitrary integer}
$i\in
%TCIMACRO{\U{2115} }%
%BeginExpansion
\mathbb{N}
%EndExpansion
$\emph{. In the same set }$\rho ^{\left( N\right) }(\mathbf{x}(t),t)$ \emph{%
is permutation-symmetric.}

$T1_{2})$ \emph{Eq.(\ref{INTEGRAL LIOUVILLE EQ}) implies that for all }$%
\left( \mathbf{x}\equiv \mathbf{x}(t),t\right) $ \emph{such that }$\mathbf{x}
$ \emph{belongs to the subset of }$\Gamma _{N}$\emph{\ in which no
interactions occur, the differential Liouville equation}%
\begin{equation}
L_{N}\left\{ \rho ^{\left( N\right) }(\mathbf{x},t)\right\} =0
\label{DIFFERENTIAL LIOUVILLE EQ}
\end{equation}%
\emph{holds, with }$L_{N}\emph{\ }\equiv \frac{\partial }{\partial t}%
+\sum\limits_{i=1,N}\mathbf{v}_{i}\cdot \nabla _{i}$ $\emph{denoting}$ $%
\emph{the}$ \emph{Liouville operator.}

$T1_{3})$ $\rho ^{\left( N\right) }(\mathbf{x},t)$ \emph{satisfies the
property of time-reversal invariance at all times }$t\in I_{E}$\emph{, where
}$I_{E}$\emph{\ is defined in Appendix A. In other words, performing a
time-reversal with respect to the fixed time origin }$t_{o}$\emph{\ the PDF
must fulfill the symmetry property}%
\begin{equation}
\emph{\ }\rho ^{\left( N\right) }(\mathbf{r},\mathbf{v},\tau +t_{o})=\rho
^{\left( N\right) }(\mathbf{r},-\mathbf{v},-\tau +t_{o}),
\label{time-reversal symmetry}
\end{equation}%
\emph{to hold for arbitrary }$t\equiv \tau +t_{o},t_{o}\in I_{E}$\emph{\ and}
$\left( \mathbf{r},\mathbf{v}\right) \equiv \left( \mathbf{r}(t),\mathbf{v}%
(t)\right) \in \overline{\Gamma }_{N}$\emph{, with }$\overline{\Gamma }_{N}$
\emph{denoting the collisionless subset of }$\Gamma _{N}$ \emph{defined in
Appendix B.}

$T1_{4})$ \emph{For an arbitrary permutation-symmetric subset }$A\subseteq
\Gamma _{N}$\emph{\ which is invariant with respect to the reversal of
colliding particle velocities, at all collision times }$t_{i}\in \left\{
t_{i}\right\} $\emph{, in validity of the Lemma it follows that }$\rho
^{\left( N\right) }(\mathbf{x},t)$\emph{\ satisfies the axiom of probability
conservation.}

\emph{Proof }- $T1_{1})$ We first notice that thanks to Axiom \#2 for all
sets $A(t_{o}),A(t)\in K(\Gamma _{N})$ it must be
\begin{equation}
\int_{A(t_{o})}d\mathbf{x}\rho ^{\left( N\right) }(\mathbf{x}%
,t)=\int_{A(t_{o})}d\mathbf{x}_{o}\rho ^{\left( N\right) }(\mathbf{x}%
_{o},t_{o}).
\end{equation}%
This property can always be restricted to permutation-symmetric sets. Then,
thanks to the identity $\left\vert \frac{\partial \mathbf{x}(t)}{\partial
\mathbf{x}_{o}}\right\vert =1$, the previous equation implies that:%
\begin{equation}
\int_{A(t_{o})}d\mathbf{x}_{o}\left[ \rho ^{\left( N\right) }(\mathbf{\chi }(%
\mathbf{x}_{o},t_{o},t),t)-\rho ^{\left( N\right) }(\mathbf{x}_{o},t_{o})%
\right] =0.  \label{PROBABILITY CONSERVATION EQUATION}
\end{equation}%
In particular, let us first require that $A(t_{o})$ and $A(t)$ correspond to
collisionless subsets, i.e., such that in the whole time interval $(t_{o},t)$
all particles of $S_{N}$ do not undergo collisions. This means that the
initial states $\mathbf{x}_{o}\in A(t_{o})$ and their images $\mathbf{x}%
\equiv \mathbf{x}(t)=\mathbf{\chi }(\mathbf{x}_{o},t_{o},t)\in A(t)$ defined
in a suitable time interval $(t_{o},t)$ are all required to be collisionless
in the same time interval$.$Then, if $\rho ^{\left( N\right) }(\mathbf{x}%
(t),t)$ is at least continuous in the time interval $(t_{o},t)$, thanks to
the arbitrariness of the set $A(t_{o})$, it follows that for any prescribed
initial state $\mathbf{x}_{o}\in A(t_{o})$ Eq.(\ref{INTEGRAL LIOUVILLE EQ})
must hold identically in the whole time interval $(t_{o},t)\subset I_{i}$
and $\rho ^{\left( N\right) }(\mathbf{x},t)$ is necessarily permutation
symmetric.

The proof of statement $T1_{2}$ is an obvious consequence of proposition $%
T1_{1}$. In fact, assuming that $\rho ^{\left( N\right) }(\mathbf{x}(t),t)$
is $C^{(k)}$, with $k\geq 1$, in the open interval $I_{i}$, differentiation
of Eq.(\ref{INTEGRAL LIOUVILLE EQ}) delivers%
\begin{equation}
\frac{d}{dt}\rho ^{\left( N\right) }(\mathbf{x}(t),t)=0,
\end{equation}%
which coincides with Eq.(\ref{DIFFERENTIAL LIOUVILLE EQ}) when identifying $%
\mathbf{x}(t)=\mathbf{x}$. Manifestly the differential Liouville equation (%
\ref{DIFFERENTIAL LIOUVILLE EQ}) holds in the subset of $\Gamma _{N}$ in
which all particles of $S_{N}$ are collisionless. Hence proposition $T1_{2}$
applies.

To prove proposition $T1_{3}$, let us perfom a time reversal with respect to
the time origin $t_{o}\in I_{E}$ of the type (\ref{TIME REVERSAL}) (see
Appendix A). Then, if $A(t_{o})$ is a collisionless subset of $\Gamma _{N}$,
we denote by $A(\tau +t_{o})$ and $A(-\tau +t_{o})$ its images at times $%
t_{1}\equiv \tau +t_{o}$ and $t_{2}\equiv -\tau +t_{o}$. For definiteness,
let us assume that the sets $A(\tau +t_{o})$ and $A(-\tau +t_{o})$ are also
collisionless subsets of $\Gamma _{N}$, while requiring that $\rho ^{\left(
N\right) }(\mathbf{x},t)$ is a stochastic PDF. Then due to Axiom \#2 and the
reversibility property (\ref{reversibility}) it must be
\begin{equation}
\int_{A(\tau +t_{o})}d\mathbf{x}_{1}\rho ^{\left( N\right) }(\mathbf{x}%
_{1},\tau +t_{o})=\int_{A(-\tau +t_{o})}d\mathbf{x}_{2}\rho ^{\left(
N\right) }(\mathbf{x}_{2},-\tau +t_{o})=\int_{A(t_{o})}d\mathbf{x}_{o}\rho
^{\left( N\right) }(\mathbf{x}_{o},t_{o}),
\end{equation}%
where $\mathbf{x}_{1}\equiv (\mathbf{r},\mathbf{v)}$ and $\mathbf{x}%
_{2}\equiv (\mathbf{r},-\mathbf{v)}$. Due to the arbitrariness of the set $%
A(t_{o})$ it follows that $\rho ^{\left( N\right) }$ necessarily must
satisfy the symmetry property (\ref{TIME REVERSAL}).

Let us now prove $T1_{4}$. First, we consider the case of single binary
collision occurring between a couple of particles $\left( l,k\right) $ of $%
S_{N}$ whose states span the $2-$body phase-space $\Gamma _{2\left(
lk\right) }\equiv \Gamma _{1(l)}\times \Gamma _{1(k)}$. For this purpose we
consider an arbitrary permutation-symmetric subset $A$ of $\Gamma _{2\left(
lk\right) }$ which is invariant with respect to the velocity transformation $%
\left( \mathbf{v}_{l},\mathbf{v}_{k}\right) \rightarrow -\left( \mathbf{v}%
_{l},\mathbf{v}_{k}\right) $, and we denote by $A^{(-)}(t_{i})$ and$\
A^{(+)}(t_{i})$ the collision subsets of $A$ corresponding to states before
and after collision events occurring at time $t_{i}$. In this case the
probabilities of the two subsets can be defined in terms of the $2-$body PDF
$\rho _{2}^{\left( N\right) }(\mathbf{x},t)$ as follows:%
\begin{equation}
\int_{A^{(-)}(t_{i})}d\mathbf{x}\rho _{2}^{\left( N\right) }(\mathbf{x}%
,t_{i})\equiv \int_{A}d\mathbf{x}_{l}d\mathbf{x}_{k}\rho _{2}^{\left(
-\right) \left( N\right) }(\mathbf{x},t_{i})\delta \left( \left\vert \mathbf{%
r}_{l}-\mathbf{r}_{k}\right\vert -\sigma \right) \Theta \left( -\mathbf{v}%
_{lk}\cdot \mathbf{n}_{lk}\right) ,  \label{nome}
\end{equation}%
\begin{equation}
\int_{A^{(+)}(t_{i})}d\mathbf{x}\rho _{2}^{\left( N\right) }(\mathbf{x}%
,t_{i})\equiv \int_{A}d\mathbf{x}_{l}d\mathbf{x}_{k}\rho _{2}^{\left(
+\right) \left( N\right) }(\mathbf{x},t_{i})\delta \left( \left\vert \mathbf{%
r}_{l}-\mathbf{r}_{k}\right\vert -\sigma \right) \Theta \left( \mathbf{v}%
_{lk}\cdot \mathbf{n}_{lk}\right) ,
\end{equation}%
where $\mathbf{v}_{lk}\equiv \mathbf{v}_{l}-\mathbf{v}_{k}$, $\mathbf{n}%
_{ij}=\mathbf{r}_{ij}/\left\vert \mathbf{r}_{ij}\right\vert $, and the two $%
\Theta -$functions in the integrals select respectively the subdomains\ of $%
A $ before and after the collision. In particular, invoking the
left-continuity of $\rho _{2}^{\left( -\right) \left( N\right) }(\mathbf{x}%
^{(-)}(t_{i}),t_{i})\equiv \rho _{2}^{\left( -\right) \left( N\right) }(%
\mathbf{x},t_{i})$, the expression (\ref{nome}) becomes%
\begin{equation}
\int_{A^{(-)}(t_{i})}d\mathbf{x}\rho _{2}^{\left( N\right) }(\mathbf{x}%
,t_{i})\equiv \int_{A}d\mathbf{x}_{l}d\mathbf{x}_{k}\rho _{2}^{\left(
N\right) }(\mathbf{x},t_{i})\delta \left( \left\vert \mathbf{r}_{l}-\mathbf{r%
}_{k}\right\vert -\sigma \right) \Theta \left( -\mathbf{v}_{lk}\cdot \mathbf{%
n}_{lk}\right) .  \label{nome-1}
\end{equation}%
Then Axiom \#2 requires that the equation%
\begin{equation}
\int_{A^{(+)}(t_{i})}d\mathbf{x}\rho _{2}^{\left( N\right) }(\mathbf{x}%
,t_{i})=\int_{A^{(-)}(t_{i})}d\mathbf{x}\rho _{2}^{\left( N\right) }(\mathbf{%
x},t_{i})  \label{PREVIOUS}
\end{equation}%
must apply identically for an arbitrary collision time $t_{i}$. Invoking the
identity%
\begin{equation}
\left\vert \frac{\partial \mathbf{x}^{(+)}(t_{i})}{\partial \mathbf{x}%
^{(-)}(t_{i})}\right\vert =1,  \label{ID-2}
\end{equation}%
which holds again at an arbitrary collision time $t_{i}\in \left\{
t_{i}\right\} $, the left-hand integral in Eq.(\ref{PREVIOUS}) can be
written explicitly as
\begin{equation}
\int_{A^{(+)}(t_{i})}d\mathbf{x}\rho _{2}^{\left( N\right) }(\mathbf{x}%
,t_{i})\equiv \int_{A}d\mathbf{r}_{l}d\mathbf{v}_{l}^{\left( +\right) }\int d%
\mathbf{r}_{k}d\mathbf{v}_{k}^{\left( +\right) }\rho _{2}^{\left( +\right)
\left( N\right) }(\mathbf{r}_{l},\mathbf{v}_{l}^{\left( +\right) },\mathbf{r}%
_{k},\mathbf{v}_{k}^{\left( +\right) },t_{i})\delta \left( \left\vert
\mathbf{r}_{l}-\mathbf{r}_{k}\right\vert -\sigma \right) \Theta \left(
\mathbf{v}_{lk}^{\left( +\right) }\cdot \mathbf{n}_{lk}\right) ,
\label{LASR}
\end{equation}%
where $\left( \mathbf{v}_{l}^{\left( +\right) },\mathbf{v}_{k}^{\left(
+\right) }\right) $ denote the particle velocities after collision. Then,
invoking the Lemma and imposing the boundary conditions according to Eq.(\ref%
{BC-2-1}), gives%
\begin{equation}
\int_{A^{(+)}(t_{i})}d\mathbf{x}\rho _{2}^{\left( N\right) }(\mathbf{x}%
,t_{i})\equiv \int_{A}d\mathbf{r}_{l}d\mathbf{v}_{l}^{\left( +\right) }\int d%
\mathbf{r}_{k}d\mathbf{v}_{k}^{\left( +\right) }\rho _{2}^{\left( N\right) }(%
\mathbf{r}_{l},\mathbf{v}_{l}^{\left( +\right) },\mathbf{r}_{k},\mathbf{v}%
_{k}^{\left( +\right) },t_{i})\delta \left( \left\vert \mathbf{r}_{l}-%
\mathbf{r}_{k}\right\vert -\sigma \right) \Theta \left( \mathbf{v}%
_{lk}^{\left( +\right) }\cdot \mathbf{n}_{lk}\right) ,
\end{equation}%
which can also be written%
\begin{equation}
\int_{A^{(+)}(t_{i})}d\mathbf{x}\rho _{2}^{\left( N\right) }(\mathbf{x}%
,t_{i})=\int_{A}d\mathbf{r}_{l}d\mathbf{v}_{l}\int d\mathbf{r}_{k}d\mathbf{v}%
_{k}\rho _{2}^{\left( N\right) }(\mathbf{r}_{l},\mathbf{v}_{l},\mathbf{v}%
_{k},\mathbf{r}_{k},t_{i})\delta \left( \left\vert \mathbf{r}_{l}-\mathbf{r}%
_{k}\right\vert -\sigma \right) \Theta \left( \mathbf{v}_{lk}\cdot \mathbf{n}%
_{lk}\right) .
\end{equation}%
Invoking assumptions 2 and 3, namely the symmetry property of the set $A$
and taking into account the time-reversal invariance of the PDF with respect
to a fixed $t_{i}$ (see previous proposition $T1_{3}$), yields%
\begin{equation}
\int_{A^{(+)}(t_{i})}d\mathbf{x}\rho _{2}^{\left( N\right) }(\mathbf{x}%
,t_{i})=\int_{A}d\mathbf{r}_{l}d\mathbf{v}_{l}\int d\mathbf{r}_{k}d\mathbf{v}%
_{k}\rho _{2}^{\left( N\right) }(\mathbf{r}_{l},\mathbf{v}_{l},\mathbf{v}%
_{k},\mathbf{r}_{k},t_{i})\delta \left( \left\vert \mathbf{r}_{l}-\mathbf{r}%
_{k}\right\vert -\sigma \right) \Theta \left( -\mathbf{v}_{lk}\cdot \mathbf{n%
}_{lk}\right) .
\end{equation}%
This expression coincides with the integral (\ref{nome-1}), thus proving the
proposition $T1_{4}$ in the case of binary collisions.

Second, let us consider the case of unary collisions. For definiteness, we
assume that particle $l$ undergoes at time $t_{i}$ a unary collision with
the boundary. In this case we can restrict to the treatment of the $1-$body
PDF $\rho _{1}^{\left( N\right) }(\mathbf{x},t)$. Then, we can identify the
sets $A^{(-)}(t_{i})$ and $A^{(+)}(t_{i})$ respectively as%
\begin{equation}
\int_{A^{(-)}(t_{i})}d\mathbf{x}\rho _{1}^{\left( N\right) }(\mathbf{x}%
,t_{i})\equiv \int_{A}d\mathbf{r}_{l}d\mathbf{v}_{l}\Theta \left( -\mathbf{v}%
_{l}\cdot \mathbf{n}_{l}\right) \delta \left( \left\vert \mathbf{r}_{l}-%
\frac{\sigma }{2}\mathbf{n}_{l}\right\vert -\frac{\sigma }{2}\right) \rho
_{1}^{\left( -\right) \left( N\right) }(\mathbf{r}_{l},\mathbf{v}_{l},t_{i}),
\end{equation}%
\begin{equation}
\int_{A^{(+)}(t_{i})}d\mathbf{x}\rho _{1}^{\left( N\right) }(\mathbf{x}%
,t_{i})\equiv \int_{A}d\mathbf{r}_{l}d\mathbf{v}_{l}^{\left( +\right)
}\Theta \left( \mathbf{v}_{l}^{\left( +\right) }\cdot \mathbf{n}_{l}\right)
\delta \left( \left\vert \mathbf{r}_{l}-\frac{\sigma }{2}\mathbf{n}%
_{l}\right\vert -\frac{\sigma }{2}\right) \rho _{1}^{\left( +\right) \left(
N\right) }(\mathbf{r}_{l},\mathbf{v}_{l}^{\left( +\right) },t_{i}),
\label{rrr}
\end{equation}%
where, again thanks to left-continuity at collision time the first integral
becomes%
\begin{equation}
\int_{A^{(-)}(t_{i})}d\mathbf{x}\rho _{1}^{\left( N\right) }(\mathbf{x}%
,t_{i})\equiv \int_{A}d\mathbf{r}_{l}d\mathbf{v}_{l}\Theta \left( -\mathbf{v}%
_{l}\cdot \mathbf{n}_{l}\right) \delta \left( \left\vert \mathbf{r}_{l}-%
\frac{\sigma }{2}\mathbf{n}_{l}\right\vert -\frac{\sigma }{2}\right) \rho
_{1}^{\left( N\right) }(\mathbf{r}_{l},\mathbf{v}_{l},t_{i}).
\end{equation}%
Taking into account the boundary conditions (\ref{BC-2-1}) for the particle $%
l$, the integral (\ref{rrr}) becomes%
\begin{equation}
\int_{A^{(+)}(t_{i})}d\mathbf{x}\rho _{1}^{\left( N\right) }(\mathbf{x}%
,t_{i})\equiv \int_{A}d\mathbf{r}_{l}d\mathbf{v}_{l}^{\left( +\right)
}\Theta \left( \mathbf{v}_{l}^{\left( +\right) }\cdot \mathbf{n}_{l}\right)
\delta \left( \left\vert \mathbf{r}_{l}-\frac{\sigma }{2}\mathbf{n}%
_{l}\right\vert -\frac{\sigma }{2}\right) \rho _{1}^{\left( N\right) }(%
\mathbf{r}_{l},\mathbf{v}_{l}^{\left( +\right) },t_{i}).
\end{equation}%
Therefore, invoking the symmetry property of the set $A$ and imposing the
time-reversal invariance for the particle $l$ with respect to a fixed
collision time for the same particle, the previous equation recovers again
Eq.(\ref{PREVIOUS}). An analogous proof can be reached in the case of
multiple collisions.

\textbf{Q.E.D.}

\bigskip

A number of remarks regarding THM.1 are in order. These include in
particular the following ones:

\begin{itemize}
\item It is important to stress that time-reversal symmetry implies that the
boundary conditions (\ref{BC-1}) can be equivalently represented by the
equation%
\begin{equation}
\rho ^{(-)\left( N\right) }(\mathbf{x}^{\left( -\right) }(t_{i}),t_{i})=\rho
^{\left( +\right) \left( N\right) }(\mathbf{x}^{(-)}(t_{i}),t_{i}).
\label{BC-11}
\end{equation}%
The two possible choices of the boundary conditions (\ref{BC-1}) and (\ref%
{BC-11}) permit one to represent respectively the outgoing or incoming PDFs,
namely $\rho ^{(+)\left( N\right) }$ and $\rho ^{(-)\left( N\right) }$, in
terms of the corresponding past and future history provided by the incoming
and outgoing PDFs respectively.

\item THM.1 proves Claim \#3. Its validity, and in particular the
proposition $T1_{3}$, can be extended also to deterministic or partially
deterministic PDFs. In particular, in the cases of the deterministic and
microcanonical $N-$body PDFs $\rho _{H}^{\left( N\right) }(\mathbf{x},t)$
and $\rho _{d}^{\left( N\right) }(\mathbf{x},t)$ (see Eqs.(\ref{DIRAC DELTA}%
) and (\ref{MICROCANONICAL PDF})), in the collisionless subsets of $\Gamma
_{N}$, Eq.(\ref{DIFFERENTIAL LIOUVILLE EQ}) implies that $\rho _{H}^{\left(
N\right) }(\mathbf{x},t)$ and $w^{\left( N\right) }(\mathbf{x},t)$ must
satisfy respectively the differential Liouville equations:%
\begin{eqnarray}
L_{N}\rho _{H}^{\left( N\right) }(\mathbf{x},t) &=&0,  \label{LIOUVILLE-1} \\
L_{N}w^{\left( N\right) }(\mathbf{x},t) &=&0.  \label{LIOUVILLE-2}
\end{eqnarray}

\item We remark that the choice provided by Eq.(\ref{BC-1}) differs in a
fundamental way from the one adopted originally by Boltzmann and
traditionally adopted in the literature (namely Eq.(\ref{PEF CONSERVATION});
see, for example, Cercignani \cite{Cercignani1988}). In fact, they coincide
if $\rho ^{\left( N\right) }(\mathbf{x},t)$ is a function only of the total
kinetic energy $E_{N}(\mathbf{x})$ (for example, a local Maxwellian PDF).
The physical interpretation in the two cases differs in a fundamental way.
In fact, Eq.(\ref{PEF CONSERVATION}) implies that, unless the $N-$body PDF$%
\rho ^{\left( N\right) }(\mathbf{x},t)$ is only a function of the total
kinetic energy $E_{N}(\mathbf{x})$, its form must change as a consequence of
arbitrary collisions. Instead, the modified boundary condition (\ref{BC-1})
requires that the functional form of the $N-$body PDF $\rho ^{\left(
N\right) }(\mathbf{x},t)$ is always preserved \emph{through arbitrary
collisions}. As a consequence, the integral Liouville equation (\ref%
{INTEGRAL LIOUVILLE EQ}) only holds for all $t_{o}$ and $t$ belonging to the
same time interval $I_{k}=\left] t_{k},t_{k+1}\right[ $ for $k\in
%TCIMACRO{\U{2115} }%
%BeginExpansion
\mathbb{N}
%EndExpansion
$ between two consecutive collision times $t_{k}<t_{k+1}$. Let us compare in
detail the consequences of the two different choices. Regarding the role of
the collision boundary condition (\ref{PEF CONSERVATION}) in the Boltzmann
equation, it is interesting to return here to the related discussion
presented by Carlo Cercignani in his last paper (Cercignani \cite%
{Cercignani2008}), where he states that: \textquotedblleft ...\textit{if it }%
[i.e., Eq.(\ref{PEF CONSERVATION})] \textit{were strictly valid at any point
of phase-space, the gain and loss terms in the Boltzmann-Grad limit would be
exactly equal. Hence there would no effect of the collisions on the time
evolution of }[$\rho _{1}^{\left( N\right) }(\mathbf{x}_{1},t)$%
]\textquotedblright . It should be stressed here that, accordingly: 1) the
factorization condition (\ref{STOSS-ANSATZ}) applies exactly only before a
binary collision occurs; 2) instead, after collision, the analogous
factorized representation holds only in an asymptotic sense. In particular,
it is always violated in a time interval after the collision which is
sufficiently close to the same event. In such a time interval the two-body
correlations would not be negligible any more. In contrast to this
viewpoint, the implication of the modified boundary condition (\ref{BC-1})
for a $2-$body PDF $\rho _{2}^{\left( N\right) }(\mathbf{x}_{1},\mathbf{x}%
_{2},t)$ satisfying before collision, in some asymptotic sense, the
factorization condition (\ref{STOSS-ANSATZ}), is that after collision the
same PDF necessarily \emph{remains factorized} in the same approximate
sense.\
\end{itemize}

\bigskip

\section{Exact H-theorems}

A \textbf{first} fundamental issue concerns the implications of the modified
boundary conditions (\ref{BC-1}) regarding the possible validity of exact
H-theorems. Here we refer in particular to the BS entropies associated
respectively with the $N-$body PDF and the corresponding $1-$body PDF. The
latter is defined as usual in terms of the $s-$body PDF (defined generally
for $s=1,N-1$) as%
\begin{equation}
\rho _{s}^{(N)}(\mathbf{x}_{1},..\mathbf{x}_{s},t)=\int\limits_{\overline{%
\Gamma }_{1(s+1)}}d\mathbf{x}_{s+1}\int\limits_{\overline{\Gamma }_{1(s+2)}}d%
\mathbf{x}_{s+2}...\int\limits_{\overline{\Gamma }_{1(N)}}d\mathbf{x}%
_{N}\rho ^{(N)}(\mathbf{x},t).  \label{REDUCED S-BODY PDF}
\end{equation}%
upon letting $s=1$, where $\overline{\Gamma }_{N}$ is the subset of the
phase-space $\Gamma _{N}$ where $\overline{\Theta }^{(N)}(\mathbf{x})=1$
(see Appendix B). In this section we prove that\ $\rho ^{\left( N\right) }$
still satisfies an exact constant H-theorem, as in the customary approach to
CSM (see Cercignani \cite{Cercignani1975}), while $\rho _{1}^{\left(
N\right) }$ admits an exact weak H-theorem. Both conditions are proved to be
a consequence of the new boundary conditions together with the microscopic
reversibility of the $S_{N}-$CDS (see Appendix A).

For definiteness, let us assume that $\rho ^{\left( N\right) }(\mathbf{x},t)$
is a stochastic PDF defined in $\Gamma _{N}\times I.$ Then necessarily $\rho
^{\left( N\right) }(\mathbf{x},t)$ manifestly must vanish in the subset of $%
\Gamma _{N}$ in which $\overline{\Theta }^{(N)}(\mathbf{x})=0$ (see Appendix
B). Then, assuming that $\rho ^{\left( N\right) }(\mathbf{x},t)$ is strictly
positive in the subset $\overline{\Gamma }_{N}$, without loss of
generalities the entropy integral (\ref{serve}) can always be restricted to
the same subset $\overline{\Gamma }_{N}$. In other words, introducing the
functional%
\begin{equation}
\overline{S}_{N}(\rho ^{\left( N\right) }(t))=-\int\limits_{\Gamma _{N}}d%
\mathbf{x}\overline{\Theta }^{(N)}(\mathbf{x})\rho ^{\left( N\right) }(%
\mathbf{x},t)\ln \rho ^{\left( N\right) }(\mathbf{x},t),
\label{BS-ENTROPY-II-}
\end{equation}%
it follows identically that%
\begin{equation}
\overline{S}_{N}(\rho ^{\left( N\right) }(t))=S_{N}(\rho ^{\left( N\right)
}(t)).  \label{ciserve}
\end{equation}%
In fact, if we replace the strong theta-function $\overline{\Theta }^{(N)}(%
\mathbf{x})$ with the weak theta-function $\Theta ^{(N)}(\mathbf{x})$ (see
Appendix B), then the two corresponding definitions for the entropy $S_{N}$
must coincide because the collision boundaries have vanishing canonical
measure.

Then the following Lemma applies.

\textbf{LEMMA to THM.2}

\emph{If }$\rho ^{\left( N\right) }(\mathbf{x},t)$ \emph{is a strictly
positive permutation-symmetric stochastic PDF,\ then the following
identities hold:}%
\begin{eqnarray}
\alpha _{1} &\equiv &\sum\limits_{i=1,N}\int\limits_{\Gamma _{N}}d\mathbf{x}%
\rho ^{\left( N\right) }(\mathbf{x},t)\ln \rho ^{\left( N\right) }(\mathbf{x}%
,t)\mathbf{v}_{i}\cdot \mathbf{n}_{i}\delta \left( \left\vert \mathbf{r}_{i}-%
\frac{\sigma }{2}\mathbf{n}_{i}\right\vert -\frac{\sigma }{2}\right) \Theta
\left( \left\vert \mathbf{r}_{i}-\mathbf{r}_{k}\right\vert -\sigma \right)
=0,  \label{id-1} \\
\alpha _{2} &\equiv &\sum\limits_{i,k=1,N;i<k}\int\limits_{\Gamma _{N}}d%
\mathbf{x}\rho ^{\left( N\right) }(\mathbf{x},t)\ln \rho ^{\left( N\right) }(%
\mathbf{x},t)\mathbf{v}_{ik}\cdot \mathbf{n}_{ik}\delta \left( \left\vert
\mathbf{r}_{i}-\mathbf{r}_{k}\right\vert -\sigma \right) \Theta \left(
\left\vert \mathbf{r}_{i}-\frac{\sigma }{2}\mathbf{n}_{i}\right\vert -\frac{%
\sigma }{2}\right) =0.  \label{id-2}
\end{eqnarray}

\emph{Proof }- Consider for example the integral (\ref{id-1}). Each term of
the sum can be decomposed as%
\begin{equation*}
\int\limits_{\Gamma _{N}}d\mathbf{xv}_{i}\cdot \mathbf{n}G=\left(
\int\limits_{\Gamma _{N}}^{\mathbf{v}_{i}\cdot \mathbf{n}_{i}>0}d\mathbf{x}%
-\int\limits_{\Gamma _{N}}^{\mathbf{v}_{i}\cdot \mathbf{n}_{i}<0}d\mathbf{x}%
\right) \left\vert \mathbf{v}_{i}\cdot \mathbf{n}_{i}\right\vert G
\end{equation*}%
where $G=\rho ^{\left( N\right) }(\mathbf{x},t)\ln \rho ^{\left( N\right) }(%
\mathbf{x},t\mathbf{)}\Theta \left( \left\vert \mathbf{r}_{i}-\mathbf{r}%
_{k}\right\vert -\sigma \right) _{i}\delta \left( \left\vert \mathbf{r}_{i}-%
\frac{\sigma }{2}\mathbf{n}_{i}\right\vert -\frac{\sigma }{2}\right) $. Now,
the two integrals on the rhs are equal in magnitude, but with opposite sign.
The proof of this proposition is analogous to that given in THM.1 (see
statement $T1_{3}$). In fact, thanks to the modified boundary conditions (%
\ref{BC-1}), the reversibility of the $S_{N}-$CDS and the symmetry property
of $\rho ^{\left( N\right) }$ stated in Axiom \#1, it follows necessarily
that%
\begin{equation}
\int\limits_{\Gamma _{N}}^{\mathbf{v}_{i}\cdot \mathbf{n}_{i}>0}d\mathbf{x}%
\left\vert \mathbf{v}_{i}\cdot \mathbf{n}_{i}\right\vert
G=\int\limits_{\Gamma _{N}}^{\mathbf{v}_{i}\cdot \mathbf{n}_{i}<0}d\mathbf{x}%
\left\vert \mathbf{v}_{i}\cdot \mathbf{n}_{i}\right\vert G.
\end{equation}%
The same type of proof applies also to the integral (\ref{id-2}).

\textbf{Q.E.D.}

\bigskip

In validity of the previous Lemma, the\emph{\ }BS entropy associated with
the $N-$body PDF\emph{\ }necessarily remains constant in time, including all
collision times. In fact, the following result applies.

\textbf{THM.2 - Constant H-theorem for the BS entropy associated with the }$%
N-$\textbf{body PDF}

\emph{Given validity of the assumptions of THM.1, we require that the }$N-$%
\emph{body PDF for }$S_{N}-$\emph{CDS }$\rho ^{\left( N\right) }(\mathbf{x}%
,t)$ \emph{is a strictly positive stochastic PDF such that its BS entropy
associated with the }$N-$\emph{body PDF }$S_{N}(\rho ^{\left( N\right) }(t))$
\emph{defined by Eq.(\ref{BS ENTROPY}) exists at }$t=t_{o}\in I$\emph{.}
\emph{Then it follows that for all }$t\in I$ \emph{the BS entropy }$%
S_{N}(\rho ^{\left( N\right) }(t))$ \emph{exists and its entropy production
rate}%
\begin{equation}
\frac{\partial }{\partial t}S_{N}(\rho ^{\left( N\right) }(t))=0
\label{EQUALITY}
\end{equation}%
\emph{vanishes identically and satisfies the constant H-theorem}%
\begin{equation}
S_{N}(\rho ^{\left( N\right) }(t))=S_{N}(\rho ^{\left( N\right) }(t_{0})).
\label{cht}
\end{equation}

\emph{Proof }- In order to prove the validity of Eq.(\ref{EQUALITY}), thanks
to the identity (\ref{ciserve}) it is sufficient to show that%
\begin{equation}
\frac{\partial }{\partial t}\overline{S}_{N}(\rho ^{\left( N\right) }(t))=0.
\end{equation}%
Hence, from Eq.(\ref{BS-ENTROPY-II-}), explicit differentiation yields%
\begin{equation}
\frac{\partial }{\partial t}\overline{S}_{N}(\rho ^{\left( N\right)
}(t))=-\int\limits_{\Gamma _{N}}d\mathbf{x}\overline{\Theta }^{(N)}(\mathbf{x%
})\frac{\partial }{\partial t}\rho ^{\left( N\right) }(\mathbf{x},t)\left[
1+\ln \rho ^{\left( N\right) }(\mathbf{x},t)\right] .
\end{equation}%
Invoking THM.1 and noting that in the collisionless subdomain in which $%
\overline{\Theta }^{(N)}(\mathbf{x})=1$ the $N-$body PDF $\rho ^{\left(
N\right) }(\mathbf{x},t)$ is by assumption differentiable and satisfies the
differential Liouville equation (\ref{DIFFERENTIAL LIOUVILLE EQ}), the
previous equation delivers:%
\begin{equation}
\frac{\partial }{\partial t}\overline{S}_{N}(\rho ^{\left( N\right)
}(t))=\sum\limits_{i=1,N}\int\limits_{\Gamma _{N}}d\mathbf{x}\overline{%
\Theta }^{(N)}\left( \mathbf{x}\right) \nabla _{i}\cdot \left[ \mathbf{v}%
_{i}\rho ^{\left( N\right) }(\mathbf{x},t)\right] \left[ 1+\ln \rho ^{\left(
N\right) }(\mathbf{x},t)\right] .
\end{equation}%
Then, Gauss theorem gives%
\begin{equation}
\frac{\partial }{\partial t}\overline{S}_{N}(\rho ^{\left( N\right)
}(t))=-\alpha _{1}-\alpha _{2},
\end{equation}%
which, thanks to the Lemma, implies Eq.(\ref{EQUALITY}). Due to the
arbitrariness of the choice of $t\in I$ it follows that $S_{N}(\rho ^{\left(
N\right) }(t))$ exists for all $t\in I$ and satisfies the constant H-theorem
(\ref{cht}).

\textbf{Q.E.D.}

\bigskip

Based on THM.2 we now proceed investigating the time evolution of the BS
entropy $S_{1}(\rho _{1}^{\left( N\right) }(t))$ associated with the $1-$%
body PDF. This leads to the proof of the validity of an exact weak H-theorem
(Claim \#4). The following result applies.

\textbf{THM.3 - Weak H-theorem for the BS entropy associated with the }$1-$%
\textbf{body PDF}

\emph{Given validity of THM.2, let us assume that }$\rho _{1}^{\left(
N\right) }(t)\equiv \rho _{1}^{\left( N\right) }(\mathbf{x}_{1},t)$\emph{\
is a strictly-positive stochastic PDF in the }$1-$\emph{body phase-space }$%
\Gamma _{1(1)}$ \emph{and define the corresponding BS entropy }$S_{1}(\rho
_{1}^{\left( N\right) }(t))$ \emph{as}%
\begin{equation}
S_{1}(\rho _{1}^{\left( N\right) }(t))=-\int\limits_{\Gamma _{1(1)}}d\mathbf{%
x}_{1}\rho _{1}^{\left( N\right) }(\mathbf{x}_{1},t)\ln \rho _{1}^{\left(
N\right) }(\mathbf{x}_{1},t).  \label{1bs}
\end{equation}%
\emph{Then it follows that for all }$t\in I$%
\begin{equation}
\frac{\partial }{\partial t}S_{1}(\rho _{1}^{\left( N\right) }(t))\geq 0
\label{WEAK H-THEOREM}
\end{equation}%
\emph{(entropy-production weak inequality).}

\emph{Proof }- To reach the proof it is sufficient to invoke the following
inequality holding for an arbitrary system of identical particles described
by strictly-positive stochastic PDFs\ $\rho ^{\left( N\right) }(\mathbf{x},t%
\mathbb{)}$ and $\rho _{1}^{\left( N\right) }(\mathbf{x}_{i},t)$, for $i=1,N$
(Brillouin Lemma \cite{Brillouin1957}):%
\begin{equation}
\int_{\Gamma _{N}}d\mathbf{x}\rho ^{\left( N\right) }(\mathbf{x},t\mathbb{)}%
\ln \rho ^{\left( N\right) }(\mathbf{x},t\mathbb{)\geq }\int_{\Gamma _{N}}d%
\mathbf{x}\rho ^{\left( N\right) }\mathbf{(x},t\mathbb{)}\ln
\prod\limits_{i=1,N}\rho _{1}^{\left( N\right) }(\mathbf{x}%
_{i},t)=N\int_{\Gamma _{1(1)}}d\mathbf{x}\rho _{1}^{\left( N\right) }\mathbf{%
(x}_{1},t\mathbb{)}\ln \rho _{1}^{\left( N\right) }(\mathbf{x}_{1},t).
\end{equation}%
This implies manifestly that%
\begin{equation}
S_{N}(\rho ^{\left( N\right) }(t))\leq NS_{1}(\rho _{1}^{\left( N\right)
}(t)),
\end{equation}%
and hence thanks to THM.2 , the inequality (\ref{WEAK H-THEOREM})
necessarily holds.

\textbf{Q.E.D.}

\bigskip

\section{Conditions of validity of Boltzmann equation and Boltzmann H-theorem%
}

Let us now investigate the conditions of validity of the Boltzmann kinetic
equation and the related H-theorem. In this context the form of the
Boltzmann equation is considered as prescribed, without dealing with its
precise construction approach. Starting point concerns the implications of
THM.1, in reference to the particular solution of the Liouville equation
realized by the deterministic $N-$body PDF $\rho _{H}^{\left( N\right) }(%
\mathbf{x},t)$. Invoking its particle factorized form\emph{\ }(\ref{DIRAC
DELTA- FACTORIZED FORM}) given in Appendix B, Eq.(\ref{LIOUVILLE-1})
manifestly requires that, for all $s=1,N$, in the collisionless subset of$\
\Gamma _{N}$ defined by the equation $\overline{\Theta }^{(N)}(\mathbf{x})=1$%
, the following hierarchy of PDEs are necessarily satisfied:%
\begin{equation}
L_{s}\prod\limits_{j=1,s}\delta \left( \mathbf{x}_{j}-\mathbf{x}%
_{j}(t)\right) =0,  \label{BBGKY hierarchy-}
\end{equation}%
with $L_{s}\equiv \frac{\partial }{\partial t}+\sum\limits_{i=1,s}\mathbf{v}%
_{i}\cdot \nabla _{i},$ while $\mathbf{x}_{i}(t)$ for $i=1,N$ denotes the $%
i- $th particle state corresponding to the initial system state $\mathbf{x}%
_{o}$ and prescribed by the $S_{N}-$CDS. Now, by the definition (\ref%
{REDUCED S-BODY PDF}), it follows that in Eq.(\ref{BBGKY hierarchy-}) the
quantity%
\begin{equation}
\rho _{Hs}^{\left( N\right) }(\mathbf{x}_{1},...\mathbf{x}_{s},t)\equiv
\prod\limits_{j=1,s}\delta \left( \mathbf{x}_{j}-\mathbf{x}_{j}(t)\right) .
\label{DETERMINISTIC
s-body PDF}
\end{equation}%
coincides with the deterministic $s-$body PDF. Therefore, Eqs.(\ref{BBGKY
hierarchy-}) are nothing but equations of the BBGKY hierarchy for $\rho
_{Hs}^{\left( N\right) }(\mathbf{x}_{1},...\mathbf{x}_{s},t)$. As a basic
consequence, in the subset of phase-space in which no collisions take place,
all these equations take the form of $s-$body Liouville equations, while the
modified boundary conditions (\ref{BC-1}) prescribe uniquely, for all $%
s=1,N-1$, the behavior of the PDFs $\rho _{Hs}^{\left( N\right) }(\mathbf{x}%
_{1},...\mathbf{x}_{s},t)$ at arbitrary collision times $\left\{
t_{i}\right\} $. Of course, the result is not surprising since in the open
time intervals $\left] t_{i},t_{i+1}\right[ $ between two arbitrary
consecutive collision events all particles of $S_{N}-$CDS behave as free
particles.

To analyze the implications of these conclusions as far as the Boltzmann
equation is concerned, let us recall the customary form of the equation
reported in the literature \cite{Boltzmann1972} and holding for a rarefied
gas \cite{Grad,Cercignani1969a,Cercignani1988}. This is given by%
\begin{equation}
L_{1}\rho _{1}^{\left( N\right) }(\mathbf{x}_{1},t)=C\left( \rho
_{2}^{\left( N\right) }\right) ,  \label{BOLTZMANN EQUATION}
\end{equation}%
where $C\left( \rho _{2}^{\left( N\right) }\right) $ denotes the Boltzmann
collision operator, namely%
\begin{equation}
C\left( \rho _{2}^{\left( N\right) }\right) \equiv K\int\limits_{U_{1(2)}}d%
\mathbf{v}_{2}\int\limits^{\mathbf{n}_{12}\cdot \mathbf{v}_{12}>0}d\Sigma
_{2}\left\vert \mathbf{n}_{12}\cdot \mathbf{v}_{12}\right\vert \left[ \rho
_{2}^{(+)\left( N\right) }-\rho _{2}^{\left( N\right) }\right] .  \label{C2}
\end{equation}%
Here the factorization (or \textquotedblleft
stosszahlansatz\textquotedblright ) conditions \cite{Boltzmann1972,Grad}
\begin{eqnarray}
\rho _{2}^{\left( N\right) } &\equiv &\rho _{1}^{\left( N\right) }(\mathbf{r}%
_{1},\mathbf{v}_{1}^{\left( -\right) },t)\rho _{1}^{\left( N\right) }(%
\mathbf{r}_{1},\mathbf{v}_{2}^{\left( -\right) },t),  \label{STOSS-ANSATZ-1}
\\
\rho _{2}^{(+)\left( N\right) } &\equiv &\rho _{1}^{\left( N\right) }(%
\mathbf{r}_{1},\mathbf{v}_{1}^{(+)},t)\rho _{1}^{\left( N\right) }(\mathbf{r}%
_{1},\mathbf{v}_{2}^{(+)},t),  \label{STOSS-ANSATZ-2}
\end{eqnarray}%
have been introduced for the incoming and outgoing $2-$body PDFs, i.e.
respectively given by Eqs.(\ref{STOSS-ANSATZ-1}) and (\ref{STOSS-ANSATZ-2}),
both evaluated at the same position $\mathbf{r}_{1}$. The rest of the
notation is standard. Thus, $K\equiv N\sigma ^{2}$, while $\int\limits^{%
\mathbf{n}_{12}\cdot \mathbf{v}_{12}>0}d\Sigma _{2}$ denotes the integration
on the subset of the solid angle element $d\Sigma _{2}$ for which $\mathbf{n}%
_{12}\cdot \mathbf{v}_{12}>0$.

Let us now address the issue of the consistency of equation (\ref{BOLTZMANN
EQUATION}) with the axiomatic approach introduced here. Based on THM.1, the
following result can be established.

\textbf{THM.4 - Conditions of validity of the Boltzmann equation}

\emph{Given validity of THM.1, the following propositions hold:}

$T4_{1}$) \emph{The form of the Boltzmann collision operator (\ref{C2})
remains unchanged if the modified boundary conditions (\ref{BC-1}) are
invoked for the factorized} $2-$\emph{body PDF }$\rho _{2}^{\left( N\right)
} $\emph{.}

$T4_{2}$) \emph{The Boltzmann equation (\ref{BOLTZMANN EQUATION}) for the }$%
1-$\emph{body PDF }$\rho _{1}^{\left( N\right) }(\mathbf{x}_{1},t)$ \emph{%
does not hold in the case }$\rho _{1}^{\left( N\right) }(\mathbf{x}_{1},t)$
\emph{is identified with} \emph{the deterministic }$1-$\emph{body\ PDF} $%
\rho _{H1}^{\left( N\right) }(\mathbf{x}_{1},t)$\emph{.}

\emph{Proof }- The proof of the first statement follows by pointing out the
identity%
\begin{equation}
\int\limits_{U_{1(2)}}d\mathbf{v}_{2}\int\limits^{\mathbf{n}_{12}\cdot
\mathbf{v}_{12}>0}d\Sigma _{2}\left\vert \mathbf{n}_{12}\cdot \mathbf{v}%
_{12}\right\vert \left[ \rho _{2}^{(+)\left( N\right) }-\rho _{2}^{\left(
N\right) }\right] =\int\limits_{U_{1(2)}}d\mathbf{v}_{2}\int\limits^{\mathbf{%
n}_{12}\cdot \mathbf{v}_{12}<0}d\Sigma _{2}\left\vert \mathbf{n}_{12}\cdot
\mathbf{v}_{12}\right\vert \left[ \rho _{2}^{(+)\left( N\right) }-\rho
_{2}^{\left( N\right) }\right] .  \label{integrale}
\end{equation}%
In fact, the integral on the solid angle element $d\Sigma _{2}$ is
manifestly not affected by the change of the orientation of the spatial
axis, which amounts to a spatial reflection whereby $\mathbf{n}%
_{12}\rightarrow -\mathbf{n}_{12}$. The rhs of Eq.(\ref{integrale}) is
consistent with the modified boundary conditions (\ref{BC-1}) to be imposed
on the factorized solution%
\begin{equation}
\rho _{2}^{\left( N\right) }\equiv \rho _{1}^{\left( N\right) }(\mathbf{r}%
_{1},\mathbf{v}_{1},t)\rho _{1}^{\left( N\right) }(\mathbf{r}_{2},\mathbf{v}%
_{2},t),
\end{equation}%
when $\mathbf{r}_{2}$ is identified with $\mathbf{r}_{1}$. We conclude that,
when Eq.(\ref{BC-2-1}) [or equivalent Eq.(\ref{BC-1})] is imposed, the form
of the Boltzmann collision operator remains unaffected.

To prove proposition $T4_{2}$) we notice that both the exact Liouville
equation (\ref{DIFFERENTIAL LIOUVILLE EQ}) and the Boltzmann kinetic
equation only hold in the same subset of the phase-space $\Gamma _{N}$ in
which collisions do not occur. This leads to a manifest contradiction.
Indeed, it is immediate to show that the Boltzmann collision operator is not
defined when the $1-$body PDF is identified with the deterministic PDF $\rho
_{H1}^{\left( N\right) }$. In fact, the product $\rho _{H1}^{\left( N\right)
}(\mathbf{r}_{1},\mathbf{v}_{1},t)\rho _{H1}^{\left( N\right) }(\mathbf{r}%
_{2},\mathbf{v}_{2},t)$ is not defined when $\mathbf{r}_{1}=\mathbf{r}_{2}$
(see Appendix B, subsection 2), namely when the supports of two
Dirac-delta's in the the product coincide.

\textbf{Q.E.D. }

\bigskip

Here the following remarks are in order. First, the reason why the modified
boundary conditions (\ref{BC-1}) do not affect the Boltzmann equation is
intrinsically due to the form of the Boltzmann collision operator. In turn,
this can be viewed as a consequence of the assumption of rarefied gas \cite%
{Grad,Cercignani1969a,Cercignani1988} for which the Boltzmann equation
applies. Second, we notice that THM.4 implies that the deterministic PDF $%
\rho _{H1}^{\left( N\right) }(t)$ cannot belong to functional subset $%
\left\{ \rho _{1}^{\left( N\right) }(t)\right\} $ to which the Boltzmann
equation applies, so that the latter can only include stochastic PDFs. As a
consequence, the Boltzmann equation does not satisfy the Axiom \#1 of the
\textquotedblleft ab initio\textquotedblright\ approach. This also proves
Claim \#5.

\bigskip

Let us now investigate the implications which concern the Boltzmann
H-theorem. For this purpose, let us assume that the BS entropy $S_{1}\left(
\rho _{1}^{\left( N\right) }(t)\right) $ associated with the $1-$body PDF $%
\rho _{1}^{\left( N\right) }(\mathbf{x}_{1},t)$ is defined. Then, according
to Boltzmann \cite{Boltzmann1972}, the theorem states that:

\begin{itemize}
\item \emph{Proposition \#1 -H-theorem:} For all $t\in I$ the entropy
production rate $\frac{\partial }{\partial t}S_{1}\left( \rho _{1}^{\left(
N\right) }(t)\right) $ is non-negative, i.e.,%
\begin{equation}
\frac{\partial }{\partial t}S_{1}\left( \rho _{1}^{\left( N\right)
}(t)\right) \geq 0.  \label{BOLTZMANN INEQUALITY}
\end{equation}

\item \emph{Proposition \#2 - H-theorem:} For all $t\in I$ the entropy
production rate vanishes identically, namely%
\begin{equation*}
\frac{\partial }{\partial t}S_{1}\left( \rho _{1}^{\left( N\right)
}(t)\right) =0,
\end{equation*}%
if and only if for all $\left( \mathbf{x},t\right) \in \Gamma _{1(1)}\times
I $, $\rho _{1}^{\left( N\right) }(t)\equiv \rho _{1}^{\left( N\right) }(%
\mathbf{x}_{1},t)$ coincides with a local $1-$body Maxwellian PDF of the form%
\begin{equation}
\rho _{1}^{\left( N\right) }(\mathbf{x}_{1},t)=\frac{1}{\pi ^{3/2}v_{th}^{2}}%
\exp \left\{ -\frac{\left( \mathbf{v}-\mathbf{V}\right) ^{2}}{v_{th}^{2}}%
\right\} ,  \label{BELOW}
\end{equation}%
where $v_{th}^{2}\left( \mathbf{r}_{1},t\right) =\frac{2T\left( \mathbf{r}%
_{1},t\right) }{m}$, $m$ is the particle mass and $\left\{ T\left( \mathbf{r}%
_{1},t\right) \geq 0,\mathbf{V}\left( \mathbf{r}_{1},t\right) \right\} $ are
suitably smooth real fluid fields.
\end{itemize}

In this reference, the following result holds.

\textbf{THM.5 - Condition of validity the Boltzmann H-theorem}

\emph{There exist particular solutions for the }$1-$\emph{body PDF }$\rho
_{1}^{\left( N\right) }(\mathbf{x}_{1},t)$ \emph{for which the Boltzmann
H-theorem (see Propositions \#1 and \#2) is violated.}

\emph{Proof} - The proof of the theorem follows thanks to the previous
considerations (see THM.4). In fact, when identifying $\rho _{1}^{\left(
N\right) }(\mathbf{x}_{1},t)\equiv \rho _{H1}^{\left( N\right) }(\mathbf{x}%
_{1},t)$, the corresponding\emph{\ }BS entropy associated with the $1-$body
PDF $S_{1}\left( \rho _{H1}^{\left( N\right) }(t)\right) $\emph{\ }vanishes
identically, i.e. (see Appendix B, subsection 3)\emph{\ }%
\begin{equation}
S_{1}\left( \rho _{H1}^{\left( N\right) }(t)\right) \equiv 0,
\label{VANISHING 1-BODY BS ENTROPY}
\end{equation}%
so that the constant H-theorem%
\begin{equation}
\frac{\partial }{\partial t}S_{1}\left( \rho _{H1}^{\left( N\right)
}(t)\right) \equiv 0  \label{constant H-theorem 1-body PDF}
\end{equation}%
necessarily holds too. Hence, $\rho _{H1}^{\left( N\right) }(x_{1},t)$
violates Proposition \#2 of the Boltzmann H-theorem.

\textbf{Q.E.D.}

The theorem proves the validity of Claim \#6. The conclusion appears
consistent with the interpretation of the BS entropy in terms of an
ignorance function on the system state. In fact, $\rho _{H1}^{\left(
N\right) }(\mathbf{x}_{1},t)$ actually generates the deterministic dynamics
of a $1-$body sub-system.

\section{Concluding remarks}

In this paper the foundations of the \textquotedblleft ab
initio\textquotedblright\ approach to the statistical description of the
Boltzmann-Sinai classical dynamical system ($S_{N}-$CDS) have been laid
down. Based on the axioms of classical statistical mechanics (CSM) and in
difference with respect to the customary treatment adopted for the Boltzmann
equation, two basic features have been pointed out:

\begin{itemize}
\item The first one is the extension of the functional class $\left\{ \rho
^{(N)}(\mathbf{x},t)\right\} $ for the $N-$body probability density to
include distributions and in particular, the deterministic $N-$body PDF $%
\rho _{H}^{(N)}(\mathbf{x},t)$ which generates the deterministic time
evolution of the underlying classical dynamical system.

\item The second one is the introduction of a new type of collision boundary
conditions which is consistent with the axiom of probability conservation
and is applicable in principle to arbitrary unary, binary or multiple
collisions. The modified boundary conditions proposed here are expected to
lead to significant new effects in the case of dense gases. Such a choice,
which differs from the one customarily adopted in the literature and
originally formulated by Boltzmann, leaves nevertheless unchanged the form
of the Boltzmann operator in the case of rarefied gases (for which the
equation actually applies).
\end{itemize}

The theoretical implications are intriguing.

In particular, it has been shown that both features are compatible with the
validity of Liouville equation (THM.1) as well as the existence of exact
H-theorems (THMs.2 and 3). As an application of the \textquotedblleft ab
initio\textquotedblright\ approach the conditions of validity of the
Boltzmann kinetic equation and related H-theorem have been investigated
(THMs.4 and 5). In contrast to the exact result provided by THM.1, it has
been shown that the deterministic $1-$body PDF $\rho _{H1}^{\left( N\right)
}(\mathbf{x}_{1},t)$ is not an admissible solution of the Boltzmann equation
(THM.4). As a consequence, the Boltzmann equation actually violates the
validity of the axioms of CSM. Similarly, in THM.5 it has been proved that
the deterministic PDF $\rho _{H1}^{\left( N\right) }(\mathbf{x}_{1},t)$ also
violates the Boltzmann H-theorem.

A further aspect concerns the original objections posed by Loschmidt and
Zermelo \cite{Loschmidt1876,Zermelo-1,Zermelo-II} and in particular their
possible relevance in the present context. This raises, in principle, the
issues of the compatibility of the \textquotedblleft ab
initio\textquotedblright\ approach with the properties of microscopic
reversibility and the Poincar\`{e} recurrence theorem,\emph{\ }both holding
for $S_{N}-$CDS. The answers are in both cases straightforward.

The solution of the first problem (Claim \#7) is provided by the weak
H-theorem pointed out in THM.3. This shows, in fact, that a non-vanishing
entropy-production rate $\frac{\partial }{\partial t}S_{1}\left( \rho
_{1}^{\left( N\right) }(t)\right) $ can occur even in the context of the
present \textquotedblleft ab initio\textquotedblright\ approach and applies
to the reversible classical dynamical system $S_{N}-$CDS (see Appendix A).
Indeed, THM.3 necessarily\emph{\ }always holds for the BS entropy associated
with the $1-$body PDF $S_{1}\left( \rho _{1}^{\left( N\right) }(t)\right) $.
Such a property applies in principle for an arbitrary stochastic $N-$body
PDF $\rho ^{\left( N\right) }(\mathbf{x}_{1},t)$ admitting the entropy
integral. Nevertheless, in contrast to the Boltzmann H-theorem, THM.3 does
not prescribe a unique form of $\rho _{1}^{\left( N\right) }(\mathbf{x}%
_{1},t)$ for which the entropy production rate vanishes identically. This
occurs, for example, if $\rho ^{\left( N\right) }(\mathbf{x},t)$ coincides
with the deterministic PDF $\rho _{H}^{\left( N\right) }(\mathbf{x},t)$.\
Therefore, it remains to be ascertained whether the constant H-theorem might
possibly be satisfied for $S_{1}\left( \rho _{1}^{\left( N\right)
}(t)\right) $ by a broader class of stochastic PDFs $\rho _{1}^{\left(
N\right) }(\mathbf{x}_{1},t)$ which determine exact solutions, rather than
asymptotic approximations, of the Liouville equation.

Analogous conclusions follow by inspecting the conditions of validity of the
Poincar\`{e} recurrence theorem in the framework of the \textquotedblleft ab
initio\textquotedblright\ approach. It is well-known that this theorem is a
characteristic property of the Boltzmann-Sinai CDS, which of course is left
unaffected in the present theory. It follows that for the same CDS the
validity of the axioms of CSM, and in particular of THMs.1-3, is warranted,
consistent with the Poincar\`{e} recurrence theorem. Finally, we notice that
the modified boundary condition introduced here is a prerequisite in order
for the deterministic $N-$body PDF to be an admissible solution of the
Liouville equation. The latter prescribes in turn, uniquely, the time
evolution of the $S_{N}-$CDS. Therefore, it follows that the modified
boundary conditions are manifestly consistent with the validity of the
Poincar\`{e} recurrence theorem (Claim \#8).

In principle, a host of further interesting questions remain to be answered.
Indeed, the results presented suggest that the adoption of the
\textquotedblleft ab initio\textquotedblright\ approach developed here
should afford a rigorous statistical description of the $S_{N}-$CDS, also
when the number and size of particles remain finite. Relevant applications
of the theory concern, therefore, in principle both dense and rarefied gases.

\section{Acknowledgments}

Part of the formal derivation of the statistical description of classical
dynamical systems presented here was described in lectures on Mathematical
Physics held by the first author at the University of Trieste in the
academic years 2009-2011.

Initial motivations for this work were based on discussions with Carlo
Cercignani (1939-2010). Further motivations have been provided by the Civil
Protection Agency of the Friuli-Venezia-Giulia (Palmanova, Udine, Italy) and
the Danieli RD Department, Danieli S.p.A. (Buttrio, Udine, Italy).

The investigation has been carried out in the framework of MIUR (Italian
Ministry for Universities and Research) PRIN Research Program
\textquotedblleft Problemi Matematici delle Teorie Cinetiche e
Applicazioni\textquotedblright , University of Trieste, Italy.

Financial support by the Italian Foundation \textquotedblleft Angelo Della
Riccia\textquotedblright\ (Firenze, Italy) is acknowledged by C.C. The same
author would also like to express his acknowledgment for the Institutional
support of Faculty of Philosophy and Science, Silesian University in Opava
(Czech Republic) and thank the project Synergy CZ.1.07/2.3.00/20.0071
sustaining the international collaboration.

\section{Appendix A:\ The Boltzmann-Sinai classical dynamical system}

In this appendix the definition and qualitative properties of the
Boltzmann-Sinai classical dynamical system $S_{N}-$CDS are recalled. For
greater clarity, the latter are first pointed out. They include in
particular the following properties:

\begin{enumerate}
\item \emph{Newtonian setting: }$S_{N}-$CDS is Newtonian, i.e., the system
state\emph{\ }$\mathbf{x\equiv }(\mathbf{r},\mathbf{v})$ belongs to an
Euclidean phase-space.

\item \emph{Domain of existence: }$S_{N}-$CDS is defined for all $t\in
I_{E}\equiv I-\left\{ t_{i}\right\} ,$ being $I\equiv
%TCIMACRO{\U{211d} }%
%BeginExpansion
\mathbb{R}
%EndExpansion
$ the real axis and $\left\{ t_{i}\right\} $ a discrete set $\left\{
t_{i}\right\} \equiv \left\{ t_{i}\in I,i\in
%TCIMACRO{\U{2115} }%
%BeginExpansion
\mathbb{N}
%EndExpansion
\right\} $ formed by the collision times (see below).

\item \emph{Microscopic autonomy: }$S_{N}-$CDS is autonomous, i.e., the
function $\mathbf{\chi }$ in Eqs.(\ref{DS})-(\ref{DS-1}) for all $t,t_{o}\in
I_{E}$ and all $\mathbf{x}_{o}\in \Gamma _{N}$ is of the form $\mathbf{\chi }%
=\mathbf{\chi }(\mathbf{x}_{o},\tau )$, with $\tau =t-t_{o}$.

\item \emph{Microscopic reversibility:} $S_{N}-$CDS is time-reversible.
Precisely, let us introduce\emph{\ }the time-reversal transformation with
respect to the time origin $t_{o}\in I_{E}$ which is defined for all $t\in
I_{E}$:%
\begin{equation}
\left\{
\begin{array}{l}
\mathbf{r}(t) \\
\mathbf{v}(t) \\
\tau \equiv t-t_{o}%
\end{array}%
\right. \rightarrow \left\{
\begin{array}{l}
\mathbf{r}^{\prime }(t^{\prime })=\mathbf{r}(t) \\
\mathbf{v}^{\prime }(t^{\prime })=-\mathbf{v}(t) \\
\tau ^{\prime }=-\tau .%
\end{array}%
\right.  \label{TIME REVERSAL}
\end{equation}%
Then, denoting $\tau =t-t_{o}$, with $t,t_{o}$ being arbitrary and $%
t,t_{o}\in I_{E}$, the phase-space function $\mathbf{\chi }$ prescribed by
the $S_{N}-$CDS [see Eqs.(\ref{DS})-(\ref{DS-1})] is such that for all $%
\mathbf{x}_{o},\mathbf{x}\equiv (\mathbf{r},\mathbf{v})\in \Gamma _{N}$ the
identity
\begin{equation}
\mathbf{x}_{o}=\mathbf{\chi }(\mathbf{r},\mathbf{v},\tau )=\mathbf{\chi }(%
\mathbf{r},-\mathbf{v},-\tau )  \label{reversibility}
\end{equation}%
holds.

\item \emph{Metric transitivity }\cite{Sinai1970,Sinai1989}\emph{: }$S_{N}-$%
CDS is metrically transitive on the energy surface, i.e., for the $S_{N}-$%
CDS the $N-$body system kinetic energy%
\begin{equation}
E_{N}(\mathbf{x})=\frac{m}{2}\sum\limits_{i=1,N}\mathbf{v}_{i}^{2},
\label{ENERGY}
\end{equation}%
is the only invariant of motion which remains constant almost everywhere on $%
\Sigma _{\alpha }$ (energy surface)%
\begin{equation}
\Sigma _{\alpha }=\left\{ \left. \mathbf{x}\right\vert E_{N}(\mathbf{x}%
)=\alpha ,\mathbf{x\in \Gamma }_{N},\alpha \in \mathbb{R}^{+}\right\} .
\label{ENERGY SURFACE}
\end{equation}
\end{enumerate}

\bigskip

Let us now introduce explicitly the dynamical system $S_{N}-$CDS. For
definiteness, we consider the ensemble $S_{N}$ of $N$ identical, smooth\ and
hard spheres of mass $m$ and diameter $\sigma $ for which:

\begin{itemize}
\item $\mathbf{x}$ is the $N-$body system state, with $\mathbf{x}\equiv
\left( \mathbf{x}_{1},..,\mathbf{x}_{N}\right) $ and $\mathbf{x}%
_{i}(t)\equiv \mathbf{x}_{i}=\left( \mathbf{r}_{i},\mathbf{v}_{i}\equiv
\frac{d\mathbf{r}_{i}(t)}{dt}\right) $ for all $i=1,N$ denotes the Newtonian
state of the $i-$th particle.

\item $\mathbf{x}(t_{o})\equiv \mathbf{x}_{o}$ is an arbitrary initial state
and $t_{o}$ an arbitrary initial time $t_{o}\in I$.

\item $\mathbf{x}_{o}$ and $\mathbf{x}$ belong to the $N-$body phase-space $%
\Gamma _{N}\equiv \Omega _{N}\times U_{N}$, where $\Omega
_{N}=\prod\limits_{i=1,N}\Omega ,$ with $\Omega \subset
%TCIMACRO{\U{211d} }%
%BeginExpansion
\mathbb{R}
%EndExpansion
^{3}$, and $U_{N}=\prod\limits_{i=1,N}U$, with $U\equiv
%TCIMACRO{\U{211d} }%
%BeginExpansion
\mathbb{R}
%EndExpansion
^{3}$, are the corresponding $N-$body and $1-$body Euclidean configuration
and velocity spaces.

Next, let us prescribe the time-evolution of $\mathbf{x}(t)\equiv \mathbf{x}$%
, i.e., the map (\ref{DS})-(\ref{DS-1}) which identifies the $S_{N}\emph{-}$%
CDS. Here we assume that all particle of $S_{N}$ can undergo only
instantaneous elastic collisions (unary, binary or multiple) occurring at
the discrete collision times $t_{i}\in \left\{ t_{i}\right\} $. In
particular, for all $i\in
%TCIMACRO{\U{2115} }%
%BeginExpansion
\mathbb{N}
%EndExpansion
$ we shall require that:
\end{itemize}

\emph{A)} \emph{The system motion} $t\rightarrow \mathbf{r}(t)$ \emph{is
inertial in all semi-open time intervals}%
\begin{equation}
I_{i}\equiv \left] t_{i},t_{i+1}\right] .  \label{OPEN INTERVAL}
\end{equation}

\emph{B) At all discrete collision times} $t_{i}\in \left\{ t_{i}\right\} $,
\emph{i.e., in which} $\mathbf{x}^{(-)}(t_{i})\equiv \mathbf{x}%
(t_{i})\rightarrow \mathbf{x}^{(+)}(t_{i})$ \emph{are collision states (see
definitions below), the }$S_{N}$\emph{-CDS\ is defined by the discrete
bijection:}%
\begin{equation}
T_{t_{i}^{(-)},t_{i}^{(+)}}:\mathbf{x}^{(-)}(t_{i})\rightarrow \mathbf{x}%
^{(+)}(t_{i})\equiv \mathbf{\chi }(\mathbf{x}%
^{(-)}(t_{i}),t_{i}^{(-)},t_{i}^{(+)})\equiv T_{t_{i}^{(-)},t_{i}^{(+)}}%
\mathbf{x}^{(-)}(t_{i}),  \label{DISCRETE DS}
\end{equation}%
\emph{with inverse}%
\begin{equation}
T_{t_{i}^{(+)},t_{i}^{(-)}}:\mathbf{x}^{(+)}(t_{i})\rightarrow \mathbf{x}%
^{(-)}(t_{i})\equiv \mathbf{\chi }(\mathbf{x}%
^{(+)}(t_{i}),t_{i}^{(+)},t_{i}^{(-)})\equiv T_{t_{i}^{(+)},t_{i}^{(-)}}%
\mathbf{x}^{(+)}(t_{i}).  \label{DISCRETE DS-1}
\end{equation}%
Here the notation is as follows. For all $t,t_{o}\in I-\left\{ t_{i}\right\}
,$ $T_{t_{o},t}$ and $T_{t,t_{o}}$ are respectively the evolution operator
and its inverse operator. Similarly, for all $t\in \left\{ t_{i}\right\} ,$ $%
T_{t_{i}^{(-)},t_{i}^{(+)}}$ and $T_{t_{i}^{(+)},t_{i}^{(-)}}$ are
respectively the corresponding discrete evolution operator and its inverse.
In addition, the vector function $\mathbf{\chi }(\mathbf{x}_{o},t_{o},t)$
introduced in Eqs.(\ref{DS})-(\ref{DS-1}) is identified with the set%
\begin{equation}
\mathbf{\chi }(\mathbf{x}_{o},t_{o},t)=\left\{ \mathbf{\chi }_{1}(\mathbf{x}%
_{o},t_{o},t),..,\mathbf{\chi }_{N}(\mathbf{x}_{o},t_{o},t)\right\} ,
\label{Chi}
\end{equation}%
where%
\begin{equation}
\mathbf{\chi }_{i}(\mathbf{x}_{o},t_{o},t)\equiv \left\{ \mathbf{\chi }_{%
\mathbf{r}_{i}}(\mathbf{x}_{o},t_{o},t),\mathbf{\chi }_{\mathbf{v}_{i}}(%
\mathbf{x}_{o},t_{o},t)\right\} ,  \label{Chi_r,Chi_v}
\end{equation}%
so that the CDS (\ref{DS}) implies also for $i=1,N$:%
\begin{eqnarray}
\mathbf{r}_{i}(t) &=&\mathbf{\chi }_{\mathbf{r}_{i}}(\mathbf{x}_{o},t_{o},t),
\label{motion application} \\
\mathbf{v}_{i}(t) &=&\mathbf{\chi }_{\mathbf{v}_{i}}(\mathbf{x}_{o},t_{o},t).
\label{velocity
application}
\end{eqnarray}%
Finally, the CDS (\ref{DS}) determines uniquely the applications%
\begin{eqnarray}
t &\rightarrow &\mathbf{r}(t)\equiv \left( \mathbf{r}_{1}(t),...\mathbf{r}%
_{N}(t)\right) ,  \label{motion} \\
t &\rightarrow &\mathbf{v}(t)\equiv \left( \mathbf{v}_{1}(t),...\mathbf{v}%
_{N}(t)\right) ,  \label{velocity} \\
t &\rightarrow &\mathbf{x}(t)\equiv \left( \mathbf{x}_{1}(t),...\mathbf{x}%
_{N}(t)\right) ,  \label{state}
\end{eqnarray}%
which are referred to respectively as motion, velocity and state applications%
\emph{\ }for $S_{N}$.

\subsection{1-Collision laws}

To define the $S_{N}-$CDS at collision time,\ we prescribe the collision
laws which apply at all the discrete collision times$\ \left\{ t_{i}\right\}
$.\ For all $t_{i}\in \left\{ t_{i}\right\} $ we denote respectively%
\begin{eqnarray}
\mathbf{x}^{\left( -\right) }(t_{i}) &\equiv &\lim_{t\rightarrow
t_{i}^{\left( -\right) }}\text{ }\mathbf{x}(t),
\label{STATE BEFORE COLLISION} \\
\mathbf{x}^{\left( +\right) }(t_{i}) &\equiv &\lim_{t\rightarrow
t_{i}^{\left( +\right) }}\text{ }\mathbf{x}(t),
\label{STATE AFTER COLLISION}
\end{eqnarray}%
as the incoming and outgoing particle states of $S_{N}$ at time $t_{i}$
(before and after collision). Notice here that by definition%
\begin{equation}
\mathbf{x}^{\left( -\right) }(t_{i})\neq \mathbf{x}^{\left( +\right)
}(t_{i}),
\end{equation}%
so that, at an arbitrary collision time $t_{i}$, the system state $\mathbf{x}%
(t)$ is left-continuous (but not right-continuous).\ Then, introducing the
discrete evolution operator%
\begin{equation}
T_{t_{i}^{(-)},t_{i}^{(+)}}\equiv \lim_{t_{a}\rightarrow t_{i}^{\left(
-\right) }}\lim_{t_{b}\rightarrow t_{i}^{\left( +\right) }}T_{t_{a},t_{b}},
\end{equation}%
and its inverse%
\begin{equation}
T_{t_{i}^{(+)},t_{i}^{(-)}}=\lim_{t_{a}\rightarrow t_{i}^{\left( -\right)
}}\lim_{t_{b}\rightarrow t_{i}^{\left( +\right) }}T_{t_{b},t_{a}},
\end{equation}%
it follows that the corresponding direct and inverse transformations become%
\begin{eqnarray}
\mathbf{x}^{\left( +\right) }(t_{i}) &=&T_{t_{i}^{(-)},t_{i}^{(+)}}\mathbf{x}%
^{\left( -\right) }(t_{i}),  \label{x-piu} \\
\mathbf{x}^{\left( -\right) }(t_{i}) &=&T_{t_{i}^{(+)},t_{i}^{(-)}}\mathbf{x}%
^{\left( +\right) }(t_{i}).  \label{x-meno-}
\end{eqnarray}%
In all cases we shall require that the motion application $t\rightarrow
\mathbf{r}(t)$ is continuous for all $t\in I$ (also for colliding
particles). Furthermore the velocity application $t\rightarrow \mathbf{v}(t)$
is left-continuous for all $t_{i}\in \left\{ t_{i}\right\} $, i.e.,%
\begin{equation}
\mathbf{v}^{\left( -\right) }(t_{i})=\mathbf{v}(t_{i}).
\end{equation}%
To determine how the operator $T_{t_{i}^{(-)},t_{i}^{(+)}}$ acts on the
particle velocities the following collision laws are assumed:

\begin{enumerate}
\item \emph{Single unary collisions }- Without loss of generality let us now
assume that: a) at a suitable collision time $t_{i}\in I$ a particle
undergoes a unary collision with the boundary $\partial \Omega $; b) the
same particle has at most a single point of contact with the same boundary;
c) the boundary $\partial \Omega $ is stationary and rigid. Then, if at time
$t_{i}$ the center of the $k-$th particle is located at position $\mathbf{r}%
_{k}(t_{i})$, its spherical surface is necessarily in contact with $\partial
\Omega $ at the position
\begin{equation}
\mathbf{r}_{k}^{\ast }(t_{i})=\mathbf{r}_{k}(t_{i})-\frac{\sigma }{2}\mathbf{%
n}_{k}(t_{i}).
\end{equation}%
Here $\mathbf{n}_{k}(t_{i})$ denotes the inward unit normal\emph{\ }vector
to the boundary $\partial \Omega $ at time $t_{i}$ and position $\mathbf{r}%
_{k}^{\ast }(t_{i})$. Then the conservation laws%
\begin{eqnarray}
&&\left. \left[ \mathbf{v}_{k}+\mathbf{v}_{k}^{(+)}\right] \cdot \mathbf{n}%
_{k}=0,\right. \\
&&\left. v_{k}^{2}=v_{k}^{(+)2},\right.
\end{eqnarray}%
must be satisfied for an elastic collision. Then, the operator $%
T_{t_{i}^{(-)},t_{i}^{(+)}}$ is defined by the unary collision law%
\begin{equation}
\mathbf{v}_{k}^{\left( +\right) }(t_{i})=\mathbf{v}_{k}(t_{i})-2\mathbf{n}%
_{k}(t_{i})\mathbf{n}_{k}(t_{i})\cdot \mathbf{v}_{k}(t_{i})\equiv
T_{t_{i}^{(-)},t_{i}^{(+)}}\mathbf{v}_{k}(t_{i}).  \label{T-3}
\end{equation}%
Notice that the collision occurs only if the incoming particle velocity is
such that $\mathbf{r}_{k}(t_{i})\cdot \mathbf{v}_{k}(t_{i})<0$. Then, for
the outgoing particle necessarily: $\mathbf{r}_{k}(t_{i})\cdot \mathbf{v}%
_{k}^{\left( +\right) }(t_{i})>0$.

\item \emph{Single binary collisions }- Let us now require that two particle
of $S_{N}$ (say particles $1$ and $2$) undergo a single elastic binary
collision at\ the collision time $t_{i}\in I$. Linear momentum and kinetic
energy conservation requires that, for $k=2$, the following two equations
are fulfilled:%
\begin{eqnarray}
\sum\limits_{j=1,k}\mathbf{v}_{j} &=&\sum\limits_{j=1,k}\mathbf{v}_{j}^{(+)},
\label{lienar momentum} \\
\sum\limits_{j=1,k}v_{j}^{2} &=&\sum\limits_{j=1,k}v_{j}^{(+)2}.
\label{kinetic energy}
\end{eqnarray}%
This implies that the particle velocities must transform as%
\begin{eqnarray}
\mathbf{v}_{1}^{\left( +\right) }(t_{i}) &=&\mathbf{v}_{1}(t)-\mathbf{n}%
_{12}(t_{i})\mathbf{n}_{12}(t_{i})\cdot \mathbf{v}_{12}(t)\equiv
T_{t_{i}^{(-)},t_{i}^{(+)}}\mathbf{v}_{1}(t_{i}),  \label{T-4} \\
\mathbf{v}_{2}^{\left( +\right) }(t_{i}) &=&\mathbf{v}_{2}(t)-\mathbf{n}%
_{21}(t_{i})\mathbf{n}_{21}(t_{i})\cdot \mathbf{v}_{21}(t)\equiv
T_{t_{i}^{(-)},t_{i}^{(+)}}\mathbf{v}_{2}(t_{i}),  \label{T-5}
\end{eqnarray}

(binary elastic collision law). In this case a necessary condition for the
binary collision to occur is that $\mathbf{n}_{12}(t_{i})\cdot \mathbf{v}%
_{12}(t_{i})<0$, while after collision: $\mathbf{n}_{12}(t_{i})\cdot \mathbf{%
v}_{12}^{\left( +\right) }(t_{i})>0$.

\item \emph{Multiple mixed (i.e., binary and/or unary) collisions }- In
multiple collisions particles are assumed to undergo at least two
simultaneous collisions. The corresponding collision laws can in principle
be determined in all cases in a straightforward way by imposing conservation
laws of the type indicated above (see Eqs.(\ref{lienar momentum}) and (\ref%
{kinetic energy})). It is important to stress that these laws are generally
independent of the single unary and binary collision laws.
\end{enumerate}

\subsection{2 - Analytic continuations}

In order that the $S_{N}-$CDS is globally defined on the time axis $I,$ $I$
being identified with the real axis $%
%TCIMACRO{\U{211d} }%
%BeginExpansion
\mathbb{R}
%EndExpansion
,$ it is necessary to introduce its analytic continuation in the set $%
\left\{ t_{i}\right\} $. This can be achieved in principle prescribing
either
\begin{equation}
\mathbf{x}(t_{i})=\mathbf{x}^{\left( -\right) }(t_{i})
\label{past-time continuation}
\end{equation}%
or
\begin{equation}
\mathbf{x}(t_{i})=\mathbf{x}^{\left( +\right) }(t_{i}),
\label{future-time continuation}
\end{equation}%
i.e., by identifying the system state at time $t_{i}$ as occurring either
before or after collision (\emph{pre}- and \emph{post-collision states}).
The definitions obtained in the two cases for $S_{N}-$CDS will be referred
to as \emph{analytic pre-} and \emph{post-collision continuations for} $%
S_{N}-$CDS. For definiteness let us denote respectively as $S_{N}^{(-)}-$CDS
and $S_{N}^{(+)}-$CDS the two CDSs. Then it follows that the corresponding
phase-space maps in Eqs.(\ref{DS})-(\ref{DS-1}), $\mathbf{\chi }^{(-)}=%
\mathbf{\chi }^{(-)}(\mathbf{x}_{o},t-t_{o})$ and $\mathbf{\chi }^{(+)}=%
\mathbf{\chi }^{(+)}(\mathbf{x}_{o},t-t_{o}),$ obtained in this way are both
globally defined for all $t,t_{o}\in I.$ \

We remark that microscopic reversibility is still preserved with such an
extended definition of the CDS only in the sense that the two continuations $%
\mathbf{\chi }^{(-)}$ and $\mathbf{\chi }^{(+)}$ are also mutually exchanged
when time-reversal is applied. In fact, consider the action of the
time-reversal tranformation with respect to an arbitrary time origin $%
t_{i}\in \left\{ t_{i}\right\} .$ Then, the states $\mathbf{x}_{o}\equiv
\mathbf{x}(t_{i})$ which for $S_{N}^{(-)}-$CDS are pre-collision states in
the initial orientation of the time-axis, become necessarily as
post-collision states in the opposite time-axis orientation and therefore
are necessarily associated with the $S_{N}^{(+)}-$CDS. In other words for $%
S_{N}^{(-)}-$CDS, $\mathbf{x}_{o}\equiv \mathbf{x}(0)=\lim_{\varepsilon
^{2}\rightarrow 0}\mathbf{x(}-\varepsilon ^{2}),$ so that $\mathbf{x}%
_{o}=\lim_{\tau \rightarrow 0^{-}}\mathbf{\chi }^{(-)}(\mathbf{r},\mathbf{v}%
,\tau ).$ Instead, after performing the time reversal, it must be $\mathbf{x}%
_{o}\equiv \mathbf{x}(0)=\lim_{\varepsilon ^{2}\rightarrow 0}\mathbf{x(}%
\varepsilon ^{2})$ which requires $\mathbf{x}_{o}$ to be considered a state
after collision, so that it must be also $\mathbf{x}_{o}\equiv \mathbf{x}%
(0)=\lim_{\tau \rightarrow 0^{+}}\mathbf{\chi }^{(+)}(\mathbf{r},\mathbf{v}%
,-\tau ).$ As a consequence, for the analytic continuations (\ref{past-time
continuation}) and (\ref{future-time continuation}), the identity (\ref%
{reversibility}) must actually be replaced with
\begin{equation}
\mathbf{x}_{o}=\mathbf{\chi }^{(-)}(\mathbf{r},\mathbf{v},\tau )=\mathbf{%
\chi }^{(+)}(\mathbf{r},-\mathbf{v},-\tau ),
\label{timepreversal conditions}
\end{equation}%
which must hold for all $\mathbf{x}_{o},\mathbf{x}\equiv (\mathbf{r},\mathbf{%
v})\in \Gamma _{N}$ and all $t,t_{o}\in I,$ with $\tau =t-t_{o}.$

\section{Appendix B: Mathematical preliminaries}

In this Appendix we set up the mathematical framework required for the
development of the \textquotedblleft ab initio\textquotedblright\ approach
to CSM.

\subsection{1-Collisionless subsets of $\Gamma _{N}$}

We first notice that for the $S_{N}-$CDS the $N-$body phase-space $\Gamma
_{N}$ can always be restricted to the subset $\overline{\Gamma }_{N}$ in
which no collisions take place (collisionless subset). Here we show that
this is the ensemble of all $\mathbf{x}\in \Gamma _{N}$ such that%
\begin{equation}
\overline{\Theta }^{(N)}(\mathbf{x})=\prod\limits_{i=1,N}\overline{\Theta }%
_{i}(\mathbf{x})\equiv \prod\limits_{i=1,N}\overline{\Theta }\left(
\left\vert \mathbf{r}_{i}-\frac{\sigma }{2}\mathbf{n}_{i}\right\vert -\frac{%
\sigma }{2}\right) \prod\limits_{j=1,i-1}\overline{\Theta }\left( \left\vert
\mathbf{r}_{i}-\mathbf{r}_{j}\right\vert -\sigma \right) =1,  \label{THETA-N}
\end{equation}%
where%
\begin{equation}
\overline{\Theta }_{i}(\mathbf{x})\equiv \overline{\Theta }\left( \left\vert
\mathbf{r}_{i}-\frac{\sigma }{2}\mathbf{n}_{i}\right\vert -\frac{\sigma }{2}%
\right) \prod\limits_{j=1,i-1}\overline{\Theta }\left( \left\vert \mathbf{r}%
_{i}-\mathbf{r}_{j}\right\vert -\sigma \right) ,  \label{I-TH THERA}
\end{equation}%
with $\overline{\Theta }(x)=\left\{
\begin{array}{lll}
1 &  & x>0 \\
0 &  & x\leq 0%
\end{array}%
\right. $ denoting the strong Heaviside theta function. Therefore, $%
\overline{\Gamma }_{N}=\prod\limits_{i=1,N}\overline{\Gamma }_{1(i)}$, where
$\overline{\Gamma }_{1(i)}=\overline{\Omega }_{1(i)}\times U_{1(i)}$ and $%
\overline{\Omega }_{1(i)}$ denotes the subset of the $i-$th particle
configuration space $\Omega _{1(i)}\subseteq
%TCIMACRO{\U{211d} }%
%BeginExpansion
\mathbb{R}
%EndExpansion
^{3}$ in which $\overline{\Theta }_{i}(\mathbf{x})=1$, namely particle $i$
does not experience unary nor binary collisions with particles in the set $%
j=1,i-1$.

\subsection{2-Deterministic, partially deterministic and stochastic PDFs}

Here we consider PDFs prescribed on $\Gamma _{N}$, $\rho ^{\left( N\right)
}(t)\equiv \rho ^{\left( N\right) }(\mathbf{x},t)$ which are defined for all
$(\mathbf{x},t)\in \Gamma _{N}\times I$ and are represented either by
ordinary functions or distributions.\ Let us first introduce the notion of
deterministic and stochastic phase-function with respect to a prescribed PDF
$\rho ^{\left( N\right) }(\mathbf{x},t)$ at time $t\in I$. If $G(\mathbf{x}%
,t)$ is a real ordinary summable function which is of class $C^{(k)}(\Gamma
_{N})$, with $k\geq 0$, and $\mathbf{x}(t)$ is the image at time $t$ of $%
\mathbf{x}(t_{o})=\mathbf{x}_{o}$ prescribed by the CDS (\ref{DS})-(\ref%
{DS-1}), then $G(\mathbf{x},t)$ is said to be deterministic (stochastic)
with respect to the PDF $\rho ^{\left( N\right) }(\mathbf{x},t)$ at time $%
t\in I$ if the identity%
\begin{equation}
G(\mathbf{x}(t),t)=\int\limits_{\Gamma _{N}}d\mathbf{x}G(\mathbf{x},t)\rho
^{\left( N\right) }(\mathbf{x},t)  \label{DETERMINISTIC G(x,t)}
\end{equation}%
holds (respectively, does not hold), with the rhs denoting the Lebesgue
integral weighted with respect to the PDF $\rho ^{\left( N\right) }(\mathbf{x%
},t)$. Let us now assume that for a suitable $\rho _{H}^{\left( N\right) }(%
\mathbf{x},t)\in \left\{ \rho ^{\left( N\right) }(t)\right\} $ the identity (%
\ref{DETERMINISTIC G(x,t)}) holds for arbitrary continuous functions $G(%
\mathbf{x},t)$, i.e.,%
\begin{equation}
G(\mathbf{x}(t),t)=\int\limits_{\Gamma _{N}}d\mathbf{x}G(\mathbf{x},t)\rho
_{H}^{\left( N\right) }(\mathbf{x},t).  \label{DETERMINISTIC CASE}
\end{equation}%
Letting in particular $G(\mathbf{x},t)=\mathbf{x}$, this requires that
identically $\mathbf{x}(t)=\int\limits_{\Gamma _{N}}d\mathbf{xx}\rho
_{H}^{\left( N\right) }(\mathbf{x},t)$, so that for all $t\in I$ the state $%
\mathbf{x}$ is deterministic with respect to $\rho _{H}^{\left( N\right) }(%
\mathbf{x},t)$. Then it follows that necessarily $\rho _{H}^{\left( N\right)
}(\mathbf{x},t)$ must coincide with the multi-dimensional Dirac delta%
\begin{equation}
\rho _{H}^{\left( N\right) }(\mathbf{x},t)=\delta \left( \mathbf{x}-\mathbf{x%
}(t)\right) ,  \label{DIRAC DELTA}
\end{equation}%
which in the following is referred to as deterministic $N-$body PDF.\emph{\ }%
Denoting by $\mathbf{x}\equiv \left( \xi _{1},..,\xi _{6N}\right) $ the
components of the system state $\mathbf{x}$ spanning $\Gamma _{N}$, in terms
of the $1-$dimensional Dirac deltas $\delta (\xi _{j}-\xi _{j}(t))$ (with $%
j=1,6N$), $\rho _{H}^{\left( N\right) }(\mathbf{x},t)$ is defined as%
\begin{equation}
\delta \left( \mathbf{x}-\mathbf{x}(t)\right) \equiv
\bigotimes\limits_{j=1,6N}\delta (\xi _{j}-\xi _{j}(t)),
\label{MULTIDIMENSIONAL DIRAC DELTA}
\end{equation}%
where $\bigotimes\limits_{j=1,6N}$ denotes the tensor product for
distributions \cite{COLOMBEAU}. Introducing the $1-$body Dirac delta's%
\begin{equation}
\rho _{H1}^{\left( N\right) }(\mathbf{x}_{i},t)\equiv \delta (\mathbf{x}_{i}-%
\mathbf{x}_{i}(t))=\delta (\mathbf{r}_{i}-\mathbf{r}_{i}(t))\delta (\mathbf{v%
}_{i}-\mathbf{v}_{i}(t)),  \label{i-th particle DIRAC DELTA}
\end{equation}%
for $i=1,N$, for systems of $N$ like-particles $\rho _{H}^{\left( N\right) }(%
\mathbf{x},t)$ can also be equivalently represented in terms of the
particle-factorized form%
\begin{equation}
\rho _{H}^{\left( N\right) }(\mathbf{x},t)=\prod\limits_{i=1,N}\rho
_{H1}^{\left( N\right) }(\mathbf{x}_{i},t)\equiv \prod\limits_{i=1,N}\delta (%
\mathbf{x}_{i}\mathbf{-x}_{i}(t)).  \label{DIRAC DELTA-
FACTORIZED FORM}
\end{equation}%
Finally, let us require that the identity (\ref{DETERMINISTIC G(x,t)}) holds
identically only for a suitable PDF $\rho _{d}^{\left( N\right) }(\mathbf{x}%
,t)\in \left\{ \rho ^{\left( N\right) }(t)\right\} $ and arbitrary
continuous functions $G(\mathbf{x},t)$ which are of the form $G=G\left\{ f(%
\mathbf{x},t)\right\} $, with $f(\mathbf{x},t)$ denoting a smooth (vector or
scalar) real function. For definiteness, let us assume that $f(\mathbf{x}%
,t)=\left\{ f_{1}(\mathbf{x},t),....f_{k}(\mathbf{x},t)\right\} $, with $%
1\leq k<6N\equiv \dim \left( \Gamma _{N}\right) $ and $f_{1}(\mathbf{x}%
,t),....f_{k}(\mathbf{x},t)$ are a set of independent smooth and
non-constant real phase-functions. This requires that for all $t\in I$ and
an arbitrary summable function $G\left\{ f(\mathbf{x},t)\right\} $ it must
be:%
\begin{equation}
G\left\{ f(\mathbf{x}(t),t)\right\} =\int\limits_{\Gamma _{N}}d\mathbf{x}%
G\left\{ f(\mathbf{x},t)\right\} \rho _{d}^{\left( N\right) }(\mathbf{x},t).
\end{equation}%
\emph{\ } Then it follows necessarily that $\rho _{d}^{\left( N\right) }(%
\mathbf{x},t)$ is of the form%
\begin{equation}
\rho _{d}^{\left( N\right) }(\mathbf{x},t)=\delta (f(\mathbf{x},t)-f(\mathbf{%
x}(t),t))w^{\left( N\right) }(\mathbf{x},t),
\label{PARTIALLY DETERMINISTIC PDF}
\end{equation}%
which is referred to as partially deterministic PDF. Here $\delta (f(\mathbf{%
x},t)-f(\mathbf{x}(t),t))$ is the $k$-dimensional Dirac delta
\begin{equation}
\delta (f(\mathbf{x},t)-f(\mathbf{x}(t),t))=\bigotimes\limits_{i=1,k}\delta
(f_{i}(\mathbf{x},t)-f_{i}(\mathbf{x}(t),t)),
\label{k-dimensional-DIRAC DELTA}
\end{equation}%
and $w^{\left( N\right) }(\mathbf{x},t)$ is a positive function independent
of $f(\mathbf{x},t)$. In particular, in Eq.(\ref{PARTIALLY DETERMINISTIC PDF}%
)\ $w^{\left( N\right) }(\mathbf{x},t)$ can always be prescribed to be a
strictly positive, smooth ordinary function. An example of a possible
realization of $\rho _{d}^{\left( N\right) }(\mathbf{x},t)$ is provided by
the PDF%
\begin{equation}
\rho _{d}^{\left( N\right) }(\mathbf{x},t)=\delta (E_{N}(\mathbf{x})-\alpha
)w^{\left( N\right) }(\mathbf{x},t),  \label{MICROCANONICAL PDF}
\end{equation}%
with $w^{\left( N\right) }(\mathbf{x},t)$ being the microcanonical $N-$body
PDF defined on the energy surface $\Sigma _{\alpha }$ and $E_{N}(\mathbf{x})$
denoting the $N-$body system kinetic energy (see Eqs.(\ref{ENERGY SURFACE})
and (\ref{DS-1})).

Finally, the PDFs $\rho ^{\left( N\right) }(\mathbf{x},t)\in \left\{ \rho
^{\left( N\right) }(t)\right\} $ which are, instead, represented by ordinary
functions will be here denoted as stochastic PDF's.\emph{\ }In the present
context it is assumed that the stochastic PDFs are also strictly positive
and suitably summable in the phase-space $\Gamma _{N}$.

Before closing this subsection, it is important to remind that the concept
of multi-dimensional Dirac delta introduced here (see in particular Eqs.(\ref%
{MULTIDIMENSIONAL DIRAC DELTA}), (\ref{i-th particle DIRAC DELTA}), (\ref%
{DIRAC DELTA- FACTORIZED FORM}) and (\ref{k-dimensional-DIRAC DELTA}))
relies on the proper definition of product between Dirac deltas. In fact,
let $f^{(a)}(x)$ and $f^{(b)}(x)$ be two real independent algebraic
functions both defined in $%
%TCIMACRO{\U{211d} }%
%BeginExpansion
\mathbb{R}
%EndExpansion
$ and with real roots respectively defined by the ensembles $\left(
x_{i}^{(a)},i=1,k^{(a)}\right) $ and $\left( x_{i}^{(b)},i=1,k^{(b)}\right) $%
. Then, the ordinary product between the two $1-$dimensional Dirac deltas,
i.e., $\delta \left( f^{(a)}(x\right) )$ and $\delta \left(
f^{(b)}(x)\right) $, can be only defined provided all the real roots $\left(
x_{i}^{(a)},i=1,k^{(a)}\right) $ differ from $\left(
x_{i}^{(b)},i=1,k^{(b)}\right) $. The result follows as an immediate
consequence from the notion of the Dirac delta as a limit function. As a
basic consequence, in the particle-factorized representation%
\begin{equation}
\rho _{H}^{\left( N\right) }(\mathbf{x},t)\equiv
\bigotimes\limits_{i=1,N}\delta (\mathbf{x}_{i}\mathbf{-x}_{i}(t))\equiv
\bigotimes\limits_{j=1,6N}\delta (\xi _{j}-\xi _{j}(t))
\end{equation}%
the components of each particle state $\mathbf{x}_{i}=(\mathbf{r}_{i},%
\mathbf{v}_{i})$, for all $i=1,N$, and hence also those of the system state $%
\xi _{j}$, for $j=1,6N$, must all be independent.\bigskip

\subsection{3-Boltzmann-Shannon entropy}

Let us now elaborate on the notion of Boltzmann-Shannon (BS) entropy to be
associated with a generic $N-$body PDF $\rho ^{\left( N\right) }(\mathbf{x}%
,t)$\ in the functional class $\left\{ \rho ^{\left( N\right) }(t)\right\} $%
. It is well-known that the Boltzmann-Shannon (BS) entropy follows from the
concept of\emph{\ }ignorance function\emph{\ }originally introduced by
Shannon in information theory (Shannon \cite{Shannon}) and further developed
by Jaynes (Jaynes \cite{Jaynes1957a,Jaynes1957b}). In the case of an
Euclidean phase-space $\Gamma _{N}$ this leads to\emph{\ }the definition%
\begin{equation}
S_{N}(\rho ^{\left( N\right) }(t))=-\int\limits_{\Gamma _{N}}d\mathbf{x}\rho
^{\left( N\right) }(\mathbf{x},t)\ln \rho ^{\left( N\right) }(\mathbf{x},t),
\label{BS ENTROPY}
\end{equation}%
which is referred to here as BS entropy associated to the $N-$body PDF. It
must be remarked that:

\begin{itemize}
\item Eq.(\ref{BS ENTROPY}) manifestly applies only when $\rho ^{\left(
N\right) }(\mathbf{x},t)$ is a strictly-positive and suitably summable
ordinary function. As a consequence it follows that the identity%
\begin{equation}
S_{N}(\rho ^{\left( N\right) }(t))=-\int\limits_{\Gamma _{N}}d\mathbf{x}%
\Theta ^{(N)}(\mathbf{x})\rho ^{\left( N\right) }(\mathbf{x},t)\ln \rho
^{\left( N\right) }(\mathbf{x},t)  \label{serve}
\end{equation}%
holds, where $\Theta ^{(N)}(\mathbf{x})=\prod\limits_{i=1,N}\Theta _{i}(%
\mathbf{x})$.

\item Eq.(\ref{BC-1}) must be properly modified in the case of a
non-Euclidean phase-space. In fact, let us consider an arbitrary phase-space
diffeomorphism $\mathbf{x}\rightarrow \mathbf{y}(\mathbf{x},t)$ mapping $%
\Gamma _{N}$ in the non-Euclidean phase-space $\Gamma _{N}^{\prime }$, and
denote\ by\ $\rho ^{\prime \left( N\right) }(\mathbf{y},t)=$\ $\rho ^{\left(
N\right) }(\mathbf{x},t)\left\vert \frac{\partial \mathbf{x}}{\partial
\mathbf{y}}\right\vert $ the corresponding PDF on $\Gamma _{N}^{\prime }$.
In order that $S_{N}(\rho ^{\left( N\right) }(t))$ remains invariant when it
is represented in terms of the transformed PDF $\rho ^{\prime \left(
N\right) }(t)\equiv \rho ^{\prime \left( N\right) }(\mathbf{y},t)$, i.e., $%
S_{N}(\rho ^{\prime \left( N\right) }(t))=S_{N}(\rho ^{\left( N\right) }(t))$%
, it follows that $S_{N}(\rho ^{\prime \left( N\right) }(t))$ must be
defined as%
\begin{equation}
S_{N}(\rho ^{\prime \left( N\right) }(t))=-\int\limits_{\Gamma _{N}^{\prime
}}d\mathbf{y}\rho ^{\prime \left( N\right) }(\mathbf{y},t)\ln \frac{\rho
^{\prime \left( N\right) }(\mathbf{y},t)}{\left\vert \frac{\partial \mathbf{x%
}}{\partial \mathbf{y}}\right\vert }.  \label{BS ENTROPY GENERAL}
\end{equation}%
This provides the recipe for the appropriate definition of the BS entropy
holding on a non-Euclidean phase-space.

\item Provided a constant H-theorem holds for $S_{N}(\rho ^{\left( N\right)
}(t))$ (see Section 2), then the BS entropy is necessarily defined for all
times $t\in I$.
\end{itemize}

A fundamental issue concerns the extension of the notion of BS entropy to
distributions of the form (\ref{PARTIALLY DETERMINISTIC PDF}). For this
purpose we notice that on the subset $\Sigma _{N,k}$ of phase-space $\Gamma
_{N}$ defined by the (vector or scalar) constraint equations $f(\mathbf{x}%
,t)-f(\mathbf{x}(t),t)=0$, with $f(\mathbf{x},t)$ denoting a smooth real
function, the PDF $\rho _{d}^{\left( N\right) }(\mathbf{x},t)$ (see Eq.(\ref%
{PARTIALLY DETERMINISTIC PDF})) actually prescribes a PDF on the same set $%
\Sigma _{N,k}$, to be identified with the reduced-dimensional PDF%
\begin{equation}
\rho ^{\left( N,k\right) }(\mathbf{x},t)=\frac{1}{\left\vert \frac{\partial f%
}{\partial \mathbf{x}}\right\vert }w^{\left( N\right) }(\mathbf{x},t).
\end{equation}%
In fact,%
\begin{equation}
1=\int\limits_{\Gamma _{N}}d\mathbf{x}\rho _{d}^{\left( N\right) }(\mathbf{x}%
,t)=\int\limits_{\Sigma _{N,k}}d\Sigma _{N,k}\frac{1}{\left\vert \frac{%
\partial f}{\partial \mathbf{x}}\right\vert }w^{\left( N\right) }(\mathbf{x}%
,t)\equiv \int\limits_{\Sigma _{N,k}}d\Sigma _{N,k}\rho ^{\left( N,k\right)
}(\mathbf{x},t).
\end{equation}%
Taking into account Eqs.(\ref{BS ENTROPY}) and (\ref{BS ENTROPY GENERAL}) it
follows that the appropriate definition of BS entropy for $\rho _{d}^{\left(
N\right) }(t)\equiv \rho _{d}^{\left( N\right) }(\mathbf{x},t)$ is
necessarily%
\begin{equation}
S_{N}(\rho _{d}^{\left( N\right) }(t))\equiv -\int\limits_{\Gamma _{N}}d%
\mathbf{x}\delta (f(\mathbf{x},t)-f(\mathbf{x}(t),t))w^{\left( N\right) }(%
\mathbf{x},t)\ln w^{\left( N\right) }(\mathbf{x},t).
\label{BS-entropy PARTIALLY DETERMINISTIC}
\end{equation}%
As a consequence, in the case of the deterministic PDF $\rho _{H}^{\left(
N\right) }(\mathbf{x},t)$, one infers that%
\begin{eqnarray}
S_{N}(\rho _{H}^{\left( N\right) }(t)) &\equiv &-\int\limits_{\Gamma _{N}}d%
\mathbf{x}\rho _{H}^{\left( N\right) }(\mathbf{x},t)\ln 1\equiv 0,
\label{ZERO N-BODY ENTROPY} \\
\frac{\partial }{\partial t}S_{N}(\rho _{H}^{\left( N\right) }(t)) &\equiv
&0,  \label{CONSTANT N-BODY ENTROPY}
\end{eqnarray}%
i.e., both the corresponding BS entropy and entropy production rate vanish
identically.


\bigskip

\bigskip

\begin{thebibliography}{99}
\bibitem{Boltzmann1972} L. Boltzmann, \textit{Weitere Studien \"{u}ber das W%
\"{a}rmegleichgewicht unter Gasmolek\"{u}len}, Wiener Berichte, \textbf{66},
275--370 (1872).

\bibitem{Loschmidt1876} J. Loschmidt, \textit{\"{U}ber den Zustand des W\"{a}%
rmegleichgewichtes eines Systemes von K\"{o}rpern mit R\"{u}cksicht auf die
Schwerkraft}, Akademie der Wissenschaften zu Wien,\textbf{73},128--142
(1876).

\bibitem{Zermelo-1} E. Zermelo, \textit{\"{U}ber einen Satz der Dynamik und
die mechanische W\"{a}rmetheorie,} Annalen der Physik \textbf{5}7, 485--494
(1896); (English translation by S.G, Brush, \textit{The kinetic theory of
gases} (pp.382--391), London, Imperial College Press 2003).

\bibitem{Zermelo-II} E. Zermelo, \textit{Uber mechanische Erkl\"{a}rungen
irreversibler Vorg\"{a}nge}, Annalen der Physik \textbf{59}, 4793--4801
(1896); (English translation by S.G. Brush, \textit{The kinetic theory of
gases} (pp. 403--411), London, Imperial College Press, 2003).

\bibitem{Boltzmann1896} L. Boltzmann, \textit{Entgegnung auf die w\"{a}%
rmetheoretischen Betrachtungen des Hrn, E. Zermelo, }Annalen der Physik
\textbf{57}, 773-784 (1896); (English translation by S.G. Brush, \textit{The
kinetic theory of gases} (pp.218-228), London, Imperial College Press 2003).

\bibitem{Boltzmann11896b} L. Boltzmann, \textit{Zu Hrn.Zermelo's Abhandlung
\"{U}ber die mechanische Erkl\textit{\"{a}}rung irreversibler Vorgange, }%
Annalen der Physik \textbf{60}, 392-398 (1897);\textit{\ }(English
translation by S.G. Brush, \textit{The kinetic theory of gases}
(pp.238-245), London, Imperial College Press 2003).

\bibitem{Boltzmann1896c} L. Boltzmann, \textit{Vorlesungen \"{u}ber
Gasstheorie},\textit{\ Vol.2, }J.A. Barth,\textit{\ }Leipzig (1896-1898);
English transl. by H. Brush, \textit{Lectures on gas theory,} University of
California Press (1964).

\bibitem{Grad} H. Grad, \textit{Thermodynamics of gases}, Handbook der
Physik \textbf{XII}, 205 (1958).

\bibitem{Lanford} O. Lanford III Jr., \textit{Time evolution of large
classical systems}, in \textit{Dynamical Systems Theory and Applications},
pp. 1-111 (E.J. Moser Ed.), Lecture Notes in Physics 38, Springer-Verlag,
Berlin (1975).

\bibitem{Cercignani1969a} C. Cercignani, \textit{Mathematical methods in
kinetic theory}, Plenum Press, New York (1969).

\bibitem{Cercignani1975} C. Cercignani, \textit{Theory and applications of
the Boltzmann equation}, Elsevier (1975).

\bibitem{Cercignani1988} C. Cercignani, \textit{The Boltzmann equation and
its applications}, Applied Mathematical Sciences 67, Springer-Verlag, Berlin
(1988).

\bibitem{Cercignani2008} C. Cercignani, \textit{134 years of Boltzmann
equation}, p. 107 in \textit{Boltzmann's legacy}, (G. Gallavotti, W. Reiter
and J. Yngvason Eds.), ESI Lecture in Mathematics and Physics, European
Mathematical Society (2008).

\bibitem{Chapman-Cowling} S. Chapman and T.G. Cowling, \textit{The
mathematical theory of non-uniform gases,} Chap.16 (Cambridge University
Press, 1939).

\bibitem{Enskog} D. Enskog, Kungl. Svensk Vetenskps Akademiens \textbf{63},
4 (1921); (English translation by S. G. Brush, \textit{Kinetic Theory}, Vol.
3, Pergamon, New York, 1972).

\bibitem{Sinai1970} Y.G Sinai Russ. Math. Surv. \textbf{25},137 (1970).

\bibitem{Sinai1989} Y.G. Sinai, \textit{Dynamical Systems II: Ergodic Theory
with Applications to Dynamical Systems and Statistical Mechanics}
(Springer-Verlag, Berlin, 1989).

\bibitem{Anosov-Sinai1967} D.V Anosov and Y.G Sinai, Russ. Math. Surv.
\textbf{22},\textbf{\ }103 (1967).

\bibitem{Shannon} C.E. Shannon, \textit{A Mathematical Theory of
Communication}, Bell System Technical Journal, vol. 27, pp. 379-423,
623-656, July, October, 1948.

\bibitem{Jaynes1957a} E. T. Jaynes, \textit{Information Theory and
Statistical Mechanics I}, Phys. Rev., \textbf{106}, 620 (1957).

\bibitem{Jaynes1957b} E. T. Jaynes, \textit{Information Theory and
Statistical Mechanics II}, Phys. Rev., \textbf{108,} 171 (1957).

\bibitem{Brillouin1957} L. Brillouin, \textit{La Science et la Th\'{e}orie
de l'Information}, (Ed. Jacques Gabay, Sceaux, 1988, facsimil first edition
Masson Ed. 1959).

%\bibitem{Cohen} E.G.D. Cohen, Physica \textbf{118A}, 17 (1983).

\bibitem{COLOMBEAU} J. F. Colombeau, \textit{Elementary introduction to new
generalized functions}, North-Holland, Amsterdam (1985).
\end{thebibliography}
\end{document}